\documentclass[useAMS,usenatbib]{mn2e}

\usepackage{multirow}
\usepackage{pifont}
\usepackage{lscape}
\usepackage{soul}

\usepackage{upgreek,amssymb}
\usepackage{amsmath}
\usepackage{multicol}
\usepackage{graphicx}
\usepackage[toc,page]{appendix}
\usepackage{enumerate}
\usepackage{xcolor}
\title[Neutron Star Kicks]{Neutron Star Kicks and their Relationship to Supernovae Ejecta Mass}
\author[J. C. Bray and J. J. Eldridge.]{J. C. Bray$^{1}$ \thanks{E-mail:
    john.bray@auckland.ac.nz}, J. J. Eldridge$^{1}$ \\ $^{1}$Department of
  Physics, University of Auckland, Private Bag 92019, Auckland, New
  Zealand} 
\pagerange{\pageref{firstpage}--\pageref{lastpage}} \pubyear{2015}

\begin{document}
\maketitle
\label{firstpage}
\begin{abstract}
We propose a simple model to explain the velocity of young neutron stars. We attempt to confirm a relationship between the amount of mass ejected in the formation of the neutron star and the `kick' velocity imparted to the compact remnant resulting from the process. We assume the velocity is given by $v_{\rm kick}=\alpha\,(M_{\rm ejecta}  / M_{\rm remnant}) + \beta\,$. To test this simple relationship we use the BPASS (Binary Population and Spectral Synthesis) code to create stellar population models from both single and binary star evolutionary pathways. We then use our Remnant Ejecta and Progenitor Explosion Relationship (REAPER) code to apply different $\alpha$ and $\beta$ values and three different `kick' orientations then record the resulting velocity probability distributions. 

We find that while a single star population provides a poor fit to the observational data, the binary population provides an excellent fit. Values of $\alpha=70\, {\rm km\,s^{-1}}$ and $\beta=110\,{\rm km\,s^{-1}}$ reproduce the \cite{RN165} observed 2-dimensional velocities and $\alpha=70\, {\rm km\,s^{-1}}$ and $\beta=120\,{\rm km\,s^{-1}}$ reproduce their inferred 3-dimensional velocity distribution for nearby single neutron stars with ages less than 3 Myrs. After testing isotropic, spin-axis aligned and orthogonal to spin-axis `kick' orientations, we find no statistical preference for a `kick' orientation. While ejecta mass cannot be the only factor that determines the velocity of supernovae compact remnants, we suggest it is a significant contributor and that the ejecta based `kick' should replace the Maxwell-Boltzmann velocity distribution currently used in many population synthesis codes.
\end{abstract}

\section{Introduction}
Neutron stars are born in the spectacular supernovae that mark the deaths of stars with initial masses above approximately 8 $M_\odot$ \citep{RN202,RN140}. The demise of these giants resulting from the gravitational collapse of the iron group core formed in the final gasp of their nuclear burning. In the process, the stellar material not retained by these exotic objects is ejected to seed the next generation of stellar offspring. 

While we know that neutron stars are born of supernovae, little has been considered about the link between the high velocity of these compact remnants, and the nature of the supernovae that generate them. The possible exception is \cite{RN239}, who proposed a link between the rotation speed of the pre-collapse cores and the neutron star `kick' velocity. An interesting feature of young neutron stars in particular, is that many are observed with high velocities, in the order of several hundred km\,s$^{-1}$ \citep[e.g.][]{RN155,RN213,RN215,RN224} with some velocities in excess of 1,000\,km\,s$^{-1}$. This is an indication that the supernovae must be highly asymmetric in some way for the neutron star to obtain such velocities. Recently \cite{RN242} found evidence for just such an asymmetry in the supernova ejecta of Cas A. This work was furthered by \cite{RN251} who successfully modelled the time evolution of the remnant using five large scale anisotropies or `pistons'. 

Using this as inspiration, we propose that the observed neutron star velocities are a result of the asymmetric ejection of the envelope and the conservation of momentum between this and the newly formed neutron star. This implies a direct relationship between the velocity of the compact remnant and the ratio of ejecta mass to neutron star mass. 

We express this mathematically as: 

\begin{equation}
\hspace{10mm}
v_{\rm kick}= \alpha\,\left(\frac{M_{\rm ejecta}}{M_{\rm remnant}}\right) + \beta
\label{equa:1}
\end{equation}
Where $v_{\rm kick}$ is the velocity, in ${\rm km\,s^{-1}}$, imparted to the neutron star as a result of the supernova, $M_{\rm ejecta}$ is the supernova ejecta mass, $M_{\rm remnant}$ is the neutron star mass and $\alpha$ and $\beta$ are unknown constants that we assume are universal for all supernovae.

The first term $\alpha$, represents the net velocity imparted to the compact remnant due to the asymmetrical ejection of the envelope. The second term $\beta$, represents the velocity contribution in the same direction from another source, possibly anisotropic neutrino emission. The purpose of introducing the $\beta$ constant is twofold, firstly it enables us to test if a constant `kick' value provides a better fit to the data than the conservation of momentum relationship proposed, and secondly it provides a `catch-all' for any other velocity contributions the compact remnant may experience.

We stress that in this work we have not sought a physical explanation of our two terms, instead our approach has been to assume a simple `kick' relationship and test its validity. Using this relationship we calculate the velocity of surviving binary systems using the equations of \cite{RN194} and, where the binary system is disrupted, we use the equations of \cite{RN146} to calculate the velocities of the individual stellar objects. 

The linking of the velocity of a neutron star to the ejecta to remnant mass ratio was explored by \cite{RN238} who considered supernovae in single star systems. In these supernovae, where the compact remnant velocity is equal to the `kick' velocity, they calculated an upper neutron star velocity limit of 500\,km\,s$^{-1}$. However, this calculation is based on the size of the turbulent structures seen in two-dimensional (2D) hydrodynamical simulations and they note that `the size of the kicks has to be postponed, until three-dimensional simulations (3D) of the whole spherical volume become feasible' \citep{RN238}. 

Further motivation to investigate asymmetric ejecta as a source of neutron star velocities comes from recent successful 3D simulation explosions. These indicate significant deformation of the shock as well as hydrodynamic instabilities such as the stalled accretion shock instability (SASI), which develop behind the stalled shock front \citep{RN294,RN307,RN306}. These processes have a duration of several hundred milliseconds and it seems feasible that they could result in large-scale asymmetries in mass densities and subsequent supernova ejecta mass. Further, \cite{RN308} suggest the rapid burning in the Si/O shells surrounding the iron core create large-scale turbulent flows resulting in significant asymmetry in the progenitor. With mounting evidence for stochastic processes that affect both the progenitor structure and the symmetry of the explosion itself, it seems appropriate to revisit asymmetric ejection of the envelope as a possible source for neutron star `kicks'. 

Advancing the work of \cite{RN238} our simulations include binary star systems where, depending on the direction of the `kick', it is possible for single neutron stars created by the disruption of the binary to gain an additional velocity from the binary orbital velocity. 

The outline of this paper is as follows. Firstly we select a subset of the pulsar observational data from \cite{RN165} to obtain the observed 2D velocities of nearby neutron stars with ages less than 3 Myr. To gain some insight into the form of the `kick', we use the maximum likelihood estimator to determine the `best fit' distribution to describe the observed 2D or tangential velocities for this neutron star subset. Secondly we outline the binary population and spectral synthesis (BPASS) code, that we use to evolve our progenitors and the remnant ejecta and progenitor explosion relationship (REAPER) code we use to create synthetic populations and probability distributions for all single star and binary system supernovae endpoints. Thirdly we generate both 2D and 3D synthetic velocity distributions for single neutron stars created by both single and binary star progenitors using different combinations of $\alpha$ and $\beta$, three different initial mass functions (IMFs), and three different `kick' orientations. We then use the Kolmogorov-Smirnov (KS) test to determine the `best fit' synthetic 2D velocity distribution to observed pulsars and the `best fit' 3D synthetic distribution to the Maxwell-Boltzmann distribution with $\sigma$=265\,km\,s$^{-1}$ obtained by \cite{RN165}. Finally we discuss our findings and present our conclusions.

\newcommand{\Myr}{\textbf{Myr}}
\section{Velocity set for neutron stars less than 3$\protect \Myr$ of age}
\label{s2}
We have compiled our observational dataset from the pulsar proper motion catalogue assembled by \cite{RN165}. This was the most comprehensive pulsar survey available containing a significant number of objects, with the advantage that the raw data is freely available. At the outset we should clarify that we are not seeking to repeat the work of \cite{RN165}, rather we are selecting a subset of their data to represent young, nearby, single neutron stars. Using this sample we hope to create an accurate representation of the natal velocity distribution for nearby neutron stars.

To ensure our subset is a representative sample, we include only those neutron stars with proper motion measurements (i.e are nearby), and that also have a characteristic age \textless3 Myr. The reason for the age limit is two-fold. Firstly, although (generally) gravitational interactions of relatively high-velocity neutron stars will have little effect on their observed velocities at later times, for low velocity stars these interactions are likely to have a more significant effect. Secondly by limiting the characteristic age we ensure our sample is not biased against high velocity neutron stars that may have travelled beyond the detection range.
 
Applying these criteria we obtain a subset of 46 neutron stars, which were then checked against the SIMBAD database \citep{RN268} to ensure no recycled objects (millisecond pulsars which may be much older) or pulsar binaries were included. One object (B2020+28), was identified as part of a disrupted pulsar-pulsar binary and was removed from the dataset \citep{RN305}. The final set is shown in Appendix 2, hereafter we will refer to this group as the `Hobbs 2D subset'. 

\subsection{Selecting the `Best fit' distribution to describe the Hobbs 2D subset}
Before we attempt to compare our neutron star subset to BPASS / REAPER synthetic velocity distributions, we first examine the  data to see if there is evidence of a single component Maxwell-Boltzmann distribution in our 2D subset. Since \cite{RN165} find a Maxwell-Boltzmann distribution as a `best fit' to their extrapolated 3D velocities, the same distribution type should be evident in the 2D observations. 

Six distribution types were selected to test, single-component Maxwell-Boltzmann distribution (1-MB), two-component mixture Maxwell-Boltzmann distribution (2-MB), single-component Gaussian distribution (1-G), two-component mixture Gaussian distribution (2-G), single-component log-normal distribution (1-LN) and two-component mixture log-normal distribution (2-LN).

To obtain a more accurate scaling factor for the two Maxwell-Boltzmann distributions we derive the probability distribution associated with a single velocity distribution comprising of the sum of two underlying velocity distributions ($V_x$ and $V_y$). (See Appendix 1).

To find the `best fit' distribution we analysed our six distributions using the maximum likelihood estimator (MLE). The MLE value was calculated by taking the logarithmic sum of each distribution's normalised probability at the 45 observed velocities. For each distribution type, we first calculate the approximate `best fit' by generating probability distributions for every combination of the variables using a step size of 10 km\,s$^{-1}$ for means and offsets and 0.1 for all other variables. The process was then repeated over a smaller region to refine our `best fit' distribution values using final step sizes of 1 km\,s$^{-1}$ for means and offsets and 0.05 for all other variables.  The MLE `best fit'  is defined as the distribution having the highest MLE value.

Our `best fits' for the six distributions tested are listed below in order of highest to lowest MLE values. In all cases $v_{xy}$ is expressed in km s$^{-1}$ and the numbers outside the brackets in the two-component distributions represent the weighting factors:\\

Two-component log-normal mixture (2-LN)
\begin{equation*}
\begin{split}
P(v_{xy} ) =0.6\left(\frac {1}{v_{xy}1.1\sqrt{2\pi}}\,\,e^{-\frac{(\log v_{xy}-\log168)^2}{2(1.1)^2}}\right)\\
+0.4\left(\frac {1}{v_{xy}0.4\sqrt{2\pi}}\,\,e^{-\frac{(\log v_{xy}-\log284)^2}{2(0.4)^2}}\right)
\end{split}
\end{equation*}
Two-component Maxwell-Boltzmann mixture (2-MB)
\begin{equation*}
P(v_{xy} ) =0.9\left(\displaystyle\frac{v_{xy}}{190^2} e^{-\frac{v_{xy}^2}{2(190)^2}}\right)\\
+0.1\left(\displaystyle\frac{v_{xy}}{786^2}\,\,e^{-\frac{v_{xy}^2}{2(786)^2}}\right)
\end{equation*}
Single-component log-normal (1-LN)
\begin{equation*}
P(v_{xy} ) =\frac {1}{v_{xy}0.95\sqrt{2\pi}}\,e^{-\frac{(\log v_{xy}-\log206)^2}{2(0.95)^2}} 
\end{equation*}
Two-component Gaussian mixture (2-G)
\begin{equation*}
\begin{split}
P(v_{xy}) =0.85\left(\displaystyle \frac{1}{136\sqrt{2\pi}}\,e^{-\frac{(v_{xy}-222)^2}{2(136)^2}}\right)\\
+0.15\left(\displaystyle\frac {1}{516\sqrt{2\pi}}\,e^{-\frac{(v_{xy}-975)^2}{2(516)^2}}\right)
\end{split}
\end{equation*}
Single-component Gaussian (1-G)
\begin{equation*}
P(v_{xy}) =\displaystyle \frac{1}{315\sqrt{2\pi}}\,e^{-\frac{(v_{xy}-307)^2}{2(315)^2}}
\end{equation*}
Single-component Maxwell-Boltzmann (1-MB)
\begin{equation*}
P(v_{xy} ) =\displaystyle \frac {v_{xy} }{311^2}\,e^{-\frac{v_{xy}^2}{2(311)^2}}
\end{equation*}\\

\begin{figure}
	\vspace{0mm}
  \centering
		\includegraphics[scale=0.70]{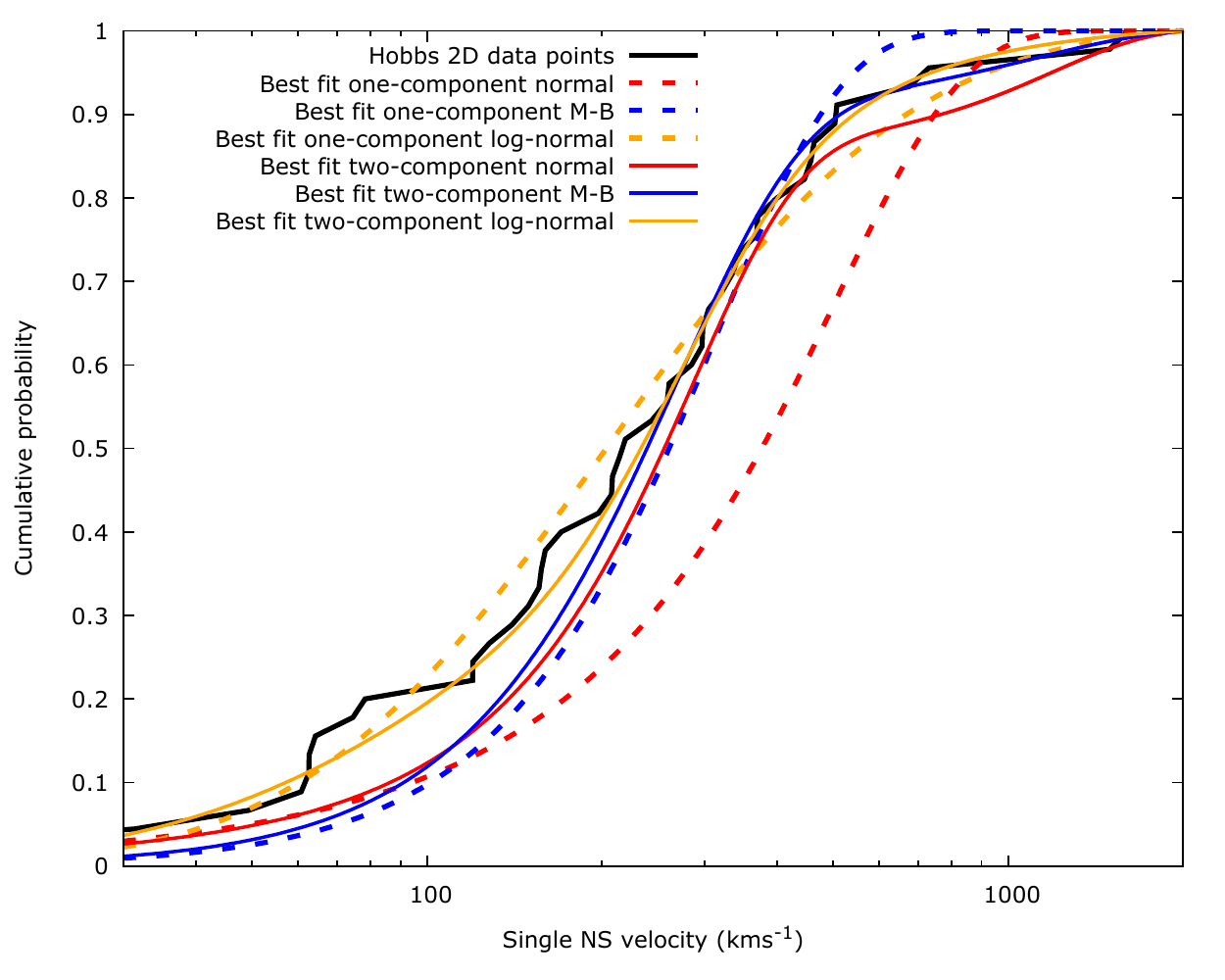}
  \caption{`Best fit' of one and two-component distributions to the Hobbs 2D subset using the maximum likelihood estimator - Cumulative Probability}
	\label{fig:1}
\end{figure}

Our `best fit' variables to the Hobbs 2D subset are shown in Figure \ref{fig:1}. We can see that, qualitatively, only the two-component log-normal mixture distribution (2-LN) provides a good fit. 

The MLE calculations showed a slight preference for the two-component mixture log-normal over the two-component mixture Maxwell-Boltzmann, single-component log-normal and two-component mixture Gaussian distributions. 

However, while our two-component log-normal mixture distribution had the highest MLE, we need some method of translating the MLE values into a statistical significance measure. To achieve this we analysed the MLE values using the Bayesian information criterion (BIC) and calculated the difference in the BIC ($\Delta$BIC) between the best fit, and each of the other five distributions. The results of this analysis are shown in Table 1 with the $\Delta$BIC column representing the difference between the most favoured distribution, the single-component log-normal distribution, and the other distributions in order of increasing $\Delta$BIC. A $\Delta$BIC of two to six indicates positive evidence for one model to be preferred over another, six to 10 indicates strong evidence while values $>10$ indicate very strong evidence.

\begin{table}	
	\caption{Results of Bayesian information criterion analysis for the six distributions tested.}
	\centering	
	\begin{tabular}{l r}
		\hline\hline
		Distribution type & $\Delta$BIC\\
		\hline\hline
		Single-component log-normal & -\\
   		Two-component Maxwell-Boltzmann mixture & 4\\
		Two-component log-normal mixture & 10\\
		Two-component Gaussian mixture & 13\\
		Single-component Gaussian & 19\\
		Single-component Maxwell-Boltzmann & 64\\
		\hline
	\end{tabular}
\end{table}

From the MLE and BIC analysis of the Hobbs 2D subset, the single-component log-normal had a positive preference over the two-component Maxwell-Boltzmann mixture and a strong preference over the two-component log-normal mixture and a very strong preference over the remaining distributions.

Mirroring our 2D analysis, recent research is also divided on whether a single-component or two-component mixture distribution is preferred to explain neutron star velocities. Applying a `deconvolution technique' to 73 pulsars aged \textless3 Myr, \cite{RN165} find a single-component Maxwell-Boltzmann distribution the `best fit' to their inferred 3D distribution. However, \cite{RN224} suggest that `some information was lost in the deconvolution process' and that the technique `may not be sensitive to subtle tail behaviour'. 

In their own analysis of isolated radio pulsars, \cite{RN224} also find a single-component velocity distribution to be the `best fit' although we note that their observational dataset contains only one pulsar with a velocity above 1,000 km\,s$^{-1}$ and the majority are below 300 km s$^{-1}$ (27 out of 34). They state their objects are `fainter and more distant than in previous proper motion surveys, reducing the bias against the high-velocity objects', but there appears to be no identification and removal of recycled objects which may be much older. This could result in an understated velocity value. 

In contrast, neutron stars in binaries provide evidence of two-component distributions, \cite{RN126} find a two-component Gaussian mixture distribution best describes the velocity distribution of low-mass x-ray binaries (LMXB's), high-mass x-ray binaries (HMXB's) and double neutron star systems, and \cite{RN215} find a two-component mixture distribution a better fit to their sample of 0.4 GHz radio pulsars. More recently, hydrodynamical simulations in 2D by \cite{RN235} show a bimodal velocity distribution and \cite{RN278} found two-component distributions evident in both spin and orbital periods of HMXB's.

The logic of a two-component velocity distribution was also explored by \cite{RN144} who proposed two-components based on the two neutrino driven collapse mechanisms, (electron capture and iron core collapse). Although, as we show below, this two-component distribution may be due to the creation of neutron stars via both single star and binary star systems.

In summary, while the MLE and BIC analysis favours a single-component log-normal distribution over the three two-component mixture distributions and the other single-component alternatives, it is worth highlighting that the $\Delta$BIC variable balances `best fit' against the number of explanatory variables and strongly penalises model complexity. Given that both single and binary star systems contribute to the neutron star population, and the wide acceptance of at least two possible collapse mechanisms within these progenitor sets, (iron core collapse and electron capture), the logic of penalising complexity may be questionable. 

\section{Numerical method}
\subsection{BPASS code}

BPASS, (\texttt{bpass.auckland.ac.nz}, \cite{RN270,RN207,RN280} and Eldridge et al., in prep) is a detailed population synthesis code, which utilises stellar models created by the Cambridge stellar evolution code STARS, originally developed by \cite{RN199}. It evolves the supernova progenitor up to the end of carbon burning and does not consider the effects of rotation or magnetic fields, but does model binary interactions in detail. While most other binary population synthesis codes use approximate stellar models, the binary models in BPASS are all evolved in a detailed stellar evolution code, which provides greater accuracy. To our knowledge, the only other code that does this is the Brussels code \citep{RN285,RN287}. 

While the BPASS code covers a range of metallicities, for our simulations we only consider progenitor sets of solar metalicity. However, there is little consensus in the literature regarding the definition of solar metallicity. \cite{RN279} and \cite{RN252}, for example, suggest the metal fraction in the Sun is close to $Z = 0.02$, while some authors, \citep{RN288,RN289} suggest that Solar metal abundances should be revised downwards to closer to $Z = 0.014$. For consistency with previous studies, we retain $Z_{\odot}=0.02$. 

We use the stellar models from BPASS v2.0 as described by \cite{RN280}. These models have been widely used {\citep{RN281,RN282,RN283}.

While a brief outline of the BPASS single star and binary evolution code is provided below, readers wishing for a more in depth understanding of the code are directed to \cite{RN270,RN207,RN280} and references therein. 

\subsubsection{Single Star Evolution}
BPASS synthesises a single star population assuming a constant star formation rate and the Kroupa initial mass function \citep{RN179}. While BPASS utilises a single star progenitor grid from 0.1 to 300 M$_\odot$, we utilise the 53 models from 5 to 150 M$_\odot$. 

 At the end of the evolution of the single star, the initial and final masses, compact remnant mass (where formed), core structure and characteristic age are written to a data file for further analysis by the REAPER code. 
 
The BPASS single star models are also used to evolve secondary star `runaways' identified by the REAPER code where the first supernova in the binary pair disrupts the system. These are shown in the `dotted' box in Appendix 3. All `runaways' are evolved as single stars with the start point for the evolution taken as the final secondary star mass after the completed evolution of the primary in the binary. `Runaways' with a final mass of $\geq8$ \,M$_\odot$  are then analysed by REAPER to determine if they experience a supernova. Those `runaways' not experiencing a supernova are assumed to remain as CO or ONeMg white dwarfs and are shown as WD1-4 in Appendix 3. 

\subsubsection{Binary Evolution}

BPASS treats the primary star as the more massive star and evolves this in detail until completion, assuming the secondary remains on the main sequence throughout the process. The primary star masses are selected from a grid that consists of star masses of 0.1, 0.2, 0.3, 0.4, 0.5, 0.6, 0.8, masses from 1 to 10 $M_\odot$ in steps of 0.5 $M_\odot$, every integer mass from 10 to 25 $M_\odot$ and then masses of 30, 35, 40, 45, 50, 60, 70, 80, 100, 120, 150, 200, and 300 $M_\odot$. In total we compute 226 single star models, 12,663 primary star models and 6,070 secondary star models, so a total of 18,959 detailed stellar evolution models. For our simulations we limit the maximum primary mass to 150 $M_\odot$. Secondary star masses are generated from the primary grid using $M_{2}/M_{1}$ ratios of 0.1, 0.2, 0.3, 0.4, 0.5, 0.6, 0.7, 0.8 and 0.9. Our probability distributions are obtained by weighting the velocities by the selected IMF of the primary star.

We assume a period distribution that is flat in log period from one day to 10$^4$ days. Recent observations suggest there may be a trend towards more close binaries \citep{RN206} but for simplicity, the flat distribution is retained and is similar to that found by \cite{RN264}. The orbits are circular and when mass is lost due to stellar wind, both spin and orbital angular momentum are reduced. The stars are assumed to rotate as a solid body, but mixing and mass loss due to rotation is ignored.

Tidal forces are ignored so stellar and orbital rotation evolve independently until Roche lobe overflow is achieved. At this point the stars are forced into a synchronous rotation with the orbit. Angular momentum is transferred between spin and orbit or vice versa. We do not include rotational mixing in our models. The rotation of the stars is only followed to allow angular momentum to be exchanged between a star's spin and the orbit.

A key advantage of BPASS is that it uses a detailed stellar evolution code to model the stars during binary interactions. However, we only model one star in detail at a time. When we evolve the primary star in detail, the secondary stars evolution is approximated by using the stellar evolution equations of \cite{RN255}. When the primary evolution is complete, its remnant mass, if a supernova occurs, is determined by calculating how much of the stellar envelope can be ejected for $10^{51}$ ergs, the typical energy in a supernova. The material remaining forms the remnant. This method is outlined in \cite{RN286} and it provides similar mass ranges for neutron stars and black hole formation as other estimates.

From these primary models REAPER is then used to calculate whether the binary would be bound or unbound in the supernova. This provides a list of binary systems in which one star remains bound to a compact object. This list of systems is then evolved in the same evolution code, but with the other star treated as a point mass compact remnant, either a white dwarf, neutron star or black hole. If the second star also experiences a supernova, REAPER again determines the fate of this system.

There are two main differences in the v2.0 BPASS stellar models compared to the earlier v1.1 models. First, rather than having separate merger models, mergers are now determined within the primary star's evolution and the mass of the companion is added to the surface of the primary star at that point. Second, the models for the binaries including a compact remnant now allow for the full range of remnants from a white dwarf of 0.1 $M_{\odot}$ up to black hole masses of 100 $M_{\odot}$. Previously only three masses were used making the treatment of compact remnant binaries less accurate.
 
\subsection{REAPER code}

REAPER is a population synthesis code written in Python\textsuperscript{\textcircled{c}} and takes the evolved BPASS models, analyses if they experience supernovae and calculates the `kicks' using Equation (1). 

For both single and binary star systems it applies the `kick' in a random direction. For binary systems, it then analyses if the system remains intact, or is disrupted, and calculates the resulting binary or component velocities. It then repeats the process for the secondary star. The complete REAPER decision tree and supernovae endpoints are shown in Appendix 3. 

The $\alpha$ and $\beta$ values are unknown and one of the aims of the REAPER code is to determine them.

In our original analysis we used the MLE to compare our 2D synthetic velocities to the Hobbs 2D subset but found this gave a very poor qualitative fit, because our comparisons only included the high $\alpha$ and $\beta$ combinations necessary to populate the highest observed velocity bins. Also, since the MLE test was not appropriate for continuous distributions, and we wished to compare the 2D and 3D results using the same test, the MLE was replaced with the KS test.
\subsubsection{Single Star Progenitors}

Our initial assumptions are as follows:
\begin{enumerate}
	\setlength\itemindent{14pt} 
	\item The single star is assumed to have a zero initial velocity. 
	\item Alignment of the single star 3D velocity to Earth is assumed to be random. 
	\item We define two criteria for a progenitor to experience a supernova. Firstly, the final evolved progenitor mass must be $\geq2 M_\odot$ and secondly the CO core mass must be $>1.4\,M_\odot$. 
	\item Following work by \cite{RN246}, the remnant is designated as a neutron star if its mass is $\leq2 M_\odot$ and as a black hole if the mass is greater than this. 
\end{enumerate}

REAPER only analyses the BPASS single star models that meet the above supernova criteria. To obtain synthetic single neutron star velocity distributions resulting from supernovae of single star progenitors, we carry out Monte Carlo simulations using a range of `kicks' obtained by cycling through all combinations of $\alpha$ and $\beta$, from 0 to 250 with a step size of 10 km\,s$^{-1}$. For each $\alpha$ and $\beta$ combination we set the 3D velocity of the remnant equal to the `kick' calculated in Equation (\ref{equa:1}). We create 3D velocity distributions by weighting each velocity by the selected IMF of the star. Following \cite{RN179} we use $dN/dM \propto M^{\Gamma}$ with $\Gamma = -2.35$. 

We then decompose each of the 3D velocities into the corresponding 2D (sky) and 1D (radial) velocities by selecting a random view angle between 0 and $\pi/2$. The process is repeated with 5,000 angles randomly chosen as follows: we select $\theta$, which represents the angle of the `kick' from the $x$ axis, by selecting a random number $x$ between 0 and 1 and calculating $\theta=\cos^{-1}(2x-1)$. For the second angle $\phi$, which represents the angle of rotation of this vector around the $x$ axis, we choose a second random number $y$ between 0 and 1 and calculate $\phi$ using $\phi=2\pi y$. In our reference frame, the $x$ axis is aligned with the orbital velocity, the $y$ axis points towards the companion and the $z$ axis is directed upward, perpendicular to both.

The resulting velocity probabilities are binned into 10 km\,s$^{-1}$ bins from 0 to 2,000 km\,s$^{-1}$, with any velocities above 2,000 km\,s$^{-1}$ added to the 2,000 km\,s$^{-1}$ bin.

To determine the `best fit' we create 2D and 3D synthetic distributions for each $\alpha$ and $\beta$ pairing, and each of our three $\Gamma$ values. The result is six velocity distributions (three 2D and three 3D), for each pairing. We then compare each 2D distribution to the Hobbs 2D subset, and each 3D distribution to the Hobbs 3D distribution using the KS test. To facilitate the comparison we use a 10 km\,s$^{-1}$ bin size for all datasets. Our 45 selected 2D data points populate 32 bin locations between 0 and 2,000 km\,s$^{-1}$ so we use $n=30$ and the KS tables to define our critical values for D$_n$. For the Hobbs 3D distribution, all 200 bin locations are populated ($n=200$) and we use the standard approximation for the critical values of 0.05$\,  =1.35/\sqrt{200}$ and 0.2$\,=1.07/\sqrt{200}$. This results in our critical values of D$_n$ for 2D and 3D shown in the keys for Table 2 and 3. Note that the 0.2 critical value represents a higher confidence level that the two samples were drawn from the same distribution.

\subsubsection{Binary Star Progenitors}
Our initial assumptions are as for the single stars, but in addition, in all velocity calculations, we ignore the effect of the shell impact on the secondary, as the effect is generally minor \citep{RN146,RN186}.

The models that experience mergers are identified within the stellar evolution model and identified as NS1 in Appendix 3. These are analysed as single stars as in section 3.2.1. 

Where there is a supernova in the primary star, REAPER separates those binaries that are disrupted and those that are intact after the supernovae. These are then analysed as below. In cases where there is no primary supernovae, REAPER treats the system as a binary intact after the primary supernovae, but with a white dwarf as the companion star instead of a compact remnant. 

For each $\alpha$ and $\beta$ combination we substitute the `kick' calculated in Equation (\ref{equa:1}) into the equations derived in \cite{RN146} and the systems as analysed as below:\\

\textbf{Binaries disrupted by the primary supernova} - The 3D spacial velocities of single neutron stars and unbound secondary stars (`runaways'), are calculated using the equations derived in \cite{RN146}. Single neutron stars, created from supernovae in primary stars that disrupt the binary system, are identified as NS2 in Appendix 3. 

The `runaways' are then evolved using the BPASS single star code. If these subsequently meet our criteria for a supernova, the compact remnant type is identified and the remnant 3D velocity gained from the secondary supernova set equal to the `kick' velocity calculated in Equation(1). Single neutron stars created by supernovae in `runaways' are identified as NS3 and NS6 in Appendix 3.

The velocities of compact remnants from supernovae in `runaways' are not necessarily aligned to the velocities received by the `runaways' from the first supernovae. To save computational time, rather than iterate over all possible `kick' directions, we assume the secondary `kick' distribution is isotropic. Therefore, the final 3D velocities for neutron stars formed by supernovae in `runaways', are obtained by multiplying the `kick' velocity gained by the remnant from the secondary supernova by $\pi/4$ and adding it at 90$^{\circ}$ to the `runaway' velocity. This $\pi/4$ factor arises from averaging over all the possible angles between the two velocities.

\textbf{Binaries intact after the primary supernovae} - If the binary remains intact after the primary supernova, we identify the remnant type and use equations in \cite{RN194} to calculate the new 3D spacial velocity of the binary system as a whole. 

We then analyse the radius of the secondary and the new system orbital period. We introduce an orbital period cut to remove those surviving systems with periods of less than 1 day, of which there are only four. Since our orbits are eccentric and BPASS' are circular, we use the new masses, separations and eccentricities to estimate the corresponding circular orbital periods. This can be assumed because, as shown by \cite{RN255}, systems with the same semi-latus rectum of their orbit evolve similarly irrespective of their eccentricity. The final primary and secondary masses and new orbital period (based on the new circular binary orbit), are binned and compared to the BPASS, secondary model set. If a match is found the secondary is evolved using BPASS assuming the remnant is a point mass. 

If the secondary is less massive than 6M$_\odot$, or the evolved secondary star does not experience a supernova according to our criteria, the system is recorded as a remnant-white dwarf binary, or in the case where there has been no primary supernova, as a white dwarf - white dwarf binary. Otherwise the secondary stars are analysed by REAPER, and where conditions for a secondary supernova are met, the compact remnant type is identified and the secondary `kick' is calculated according to Equation (1). Once again we use equations derived in \cite{RN146} to determine if the binary remains intact or is unbound by the secondary supernova.

Where our analysis shows the binary is still intact after the secondary supernova, the 3D spacial velocity of the remnant-remnant, or white dwarf - remnant system, is calculated using equations derived in \cite{RN194}. 

Where the binary is disrupted by the secondary supernova, the remnant type is identified, and the velocities of the remnant and unbound secondary are calculated using the equations derived in \cite{RN146}. In the case where there has been no primary supernova, the initial velocity of the system is assumed to be zero. 

Where there has been a primary supernova, the velocities resulting from the secondary supernovae are again, not necessarily aligned to the velocity resulting from the first supernova. Hence we calculate our final surviving system or stellar component velocities by adding the average velocities from the secondary supernova at 90$^{\circ}$ to the initial system velocity as before. Single neutron stars, created as a result of supernovae in secondary stars disrupting the binary systems, are identified in Appendix 3 as NS8, if there was no primary supernova, and as NS5 and NS7 if the binary survived the primary supernova. Note that NS4 is a bound neutron star created by the primary supernova and subsequently unbound by the secondary supernovae. This neutron star is excluded from our analysis due to its age being $>$3 Myrs.

In each instance, we repeat the process using 5,000 randomly chosen angles as described in section 3.2.1. We then decompose each of the 3D velocities into the corresponding 2D (sky) and 1D (radial) velocities as before. We then bin all the 3D and 2D velocity probability sets into 10 km\,s$^{-1}$ bins from 0 to 2,000 km\,s$^{-1}$. In all cases any velocities recorded above 2,000 km\,s$^{-1}$ are added to the 2,000 km\,s$^{-1}$ bin. We then proceed as per single star progenitors to determine our `best fit' combination using the KS test.
 
\subsubsection{Varied model parameters and caveats}

We explored some of the variations and caveats in our population synthesis, which we detail here.\\
\\
\begin{enumerate}
\setlength\itemindent{14pt} 
\item Our present analysis does not include supernova reversal, where the secondary accretes enough material so that it's evolution is accelerated and it explodes before the primary. At solar metallicity this is rare, as discussed by \cite{RN207}. The code does however, allow the evolution and analysis of the secondary star when there is no primary supernovae so we can obtain white dwarf-neutron star (WD-NS), and white dwarf-black hole (WD-BH), binaries via this pathway.
\item In addition to the \cite{RN179} IMF, we follow \cite{RN284} and test $\Gamma \pm$ 0.35 for both our 2D and 3D distributions. This results in six velocity distributions for each $\alpha$ and $\beta$ pairing.
\item To test the robustness of the `kick' calculations, we repeat the process using only an $\alpha$ component and then only a $\beta$ component.
\item For the first supernovae only, in the binary we also vary the `kick' alignment. In addition to our isotropic `kick', following work by \cite{RN203} we also test a polar (spin axis aligned) `kick' with a maximum deviation of $\pm$ 30$^{\circ}$ around the positive and negative z-axes, and an equatorial (orthogonal to the spin axis) `kick' with a maximum deviation of $\pm$ 10$^{\circ}$ around the x-y plane. 
For both the polar and equatorial `kicks', because the coordinate system used by \cite{RN146} is centred around the x-axis, we create distribution in these bands around the x-axis and then rotate the points to be distributed around the polar and equatorial regions of the z-axis. 
\item Where the secondary in a bound binary experiences a supernova, the secondary `kick' is assumed isotropic and the velocities obtained by the intact binary, or unbound individual components, are averaged and added at 90$^{\circ}$ to the original 3D system velocity.
\item We do not consider electron-capture supernovae.
\end{enumerate}
Our results are not expected to be substantially altered by these omissions and approximations. In point (i), supernovae reversal are relatively rare especially since in our models, the primary is larger than the secondary. In the case of point (iv), we show in Table 3 that the `kick' orientation has little effect on the `best fit' variables, and lastly in point (vi), electron capture supernovae only represent approximately 4 per cent of single star supernovae \citep{RN144,RN204,RN290}. Further, recent studies \citep{RN309,RN310} have show that iron core collapse supernovae in binaries can also occur with very low ejecta masses, of the order of 0.1 M$_\odot$.  We find the number of such systems in our models would be less than a few percent, and that these systems, and the electron capture supernovae, would only result in small `kicks', which are unlikely to unbind the binary systems, and therefore not produce single neutron stars.

\begin{figure}
		\includegraphics[scale=0.70]{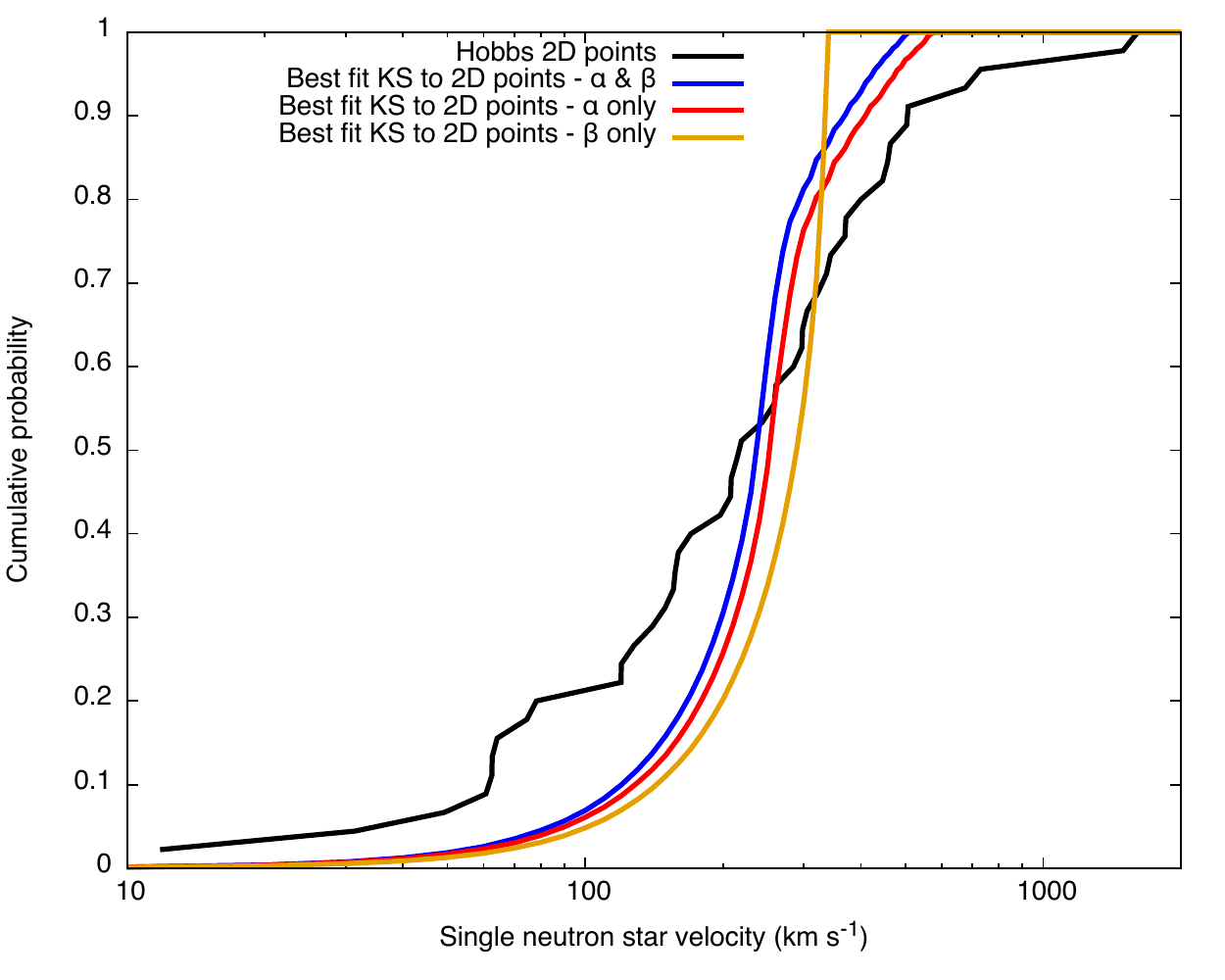}
	\caption{Single stars - `best fit' synthetic two-dimensional (2D) velocity distributions to the Hobbs 2D subset using the Kolmogorov-Smirnov test - Cumulative probability for $\Gamma$=-2.35, KS `best fit' : $\alpha$ and $\beta$ ($\alpha=50$, $\beta=30$), $\alpha$ only ($\alpha=60$), $\beta$ only ($\beta=340$)}
	\label{fig:2}
\end{figure}
\begin{figure}
	\vspace{0mm}
	\hspace{0mm}
		\includegraphics[scale=0.70]{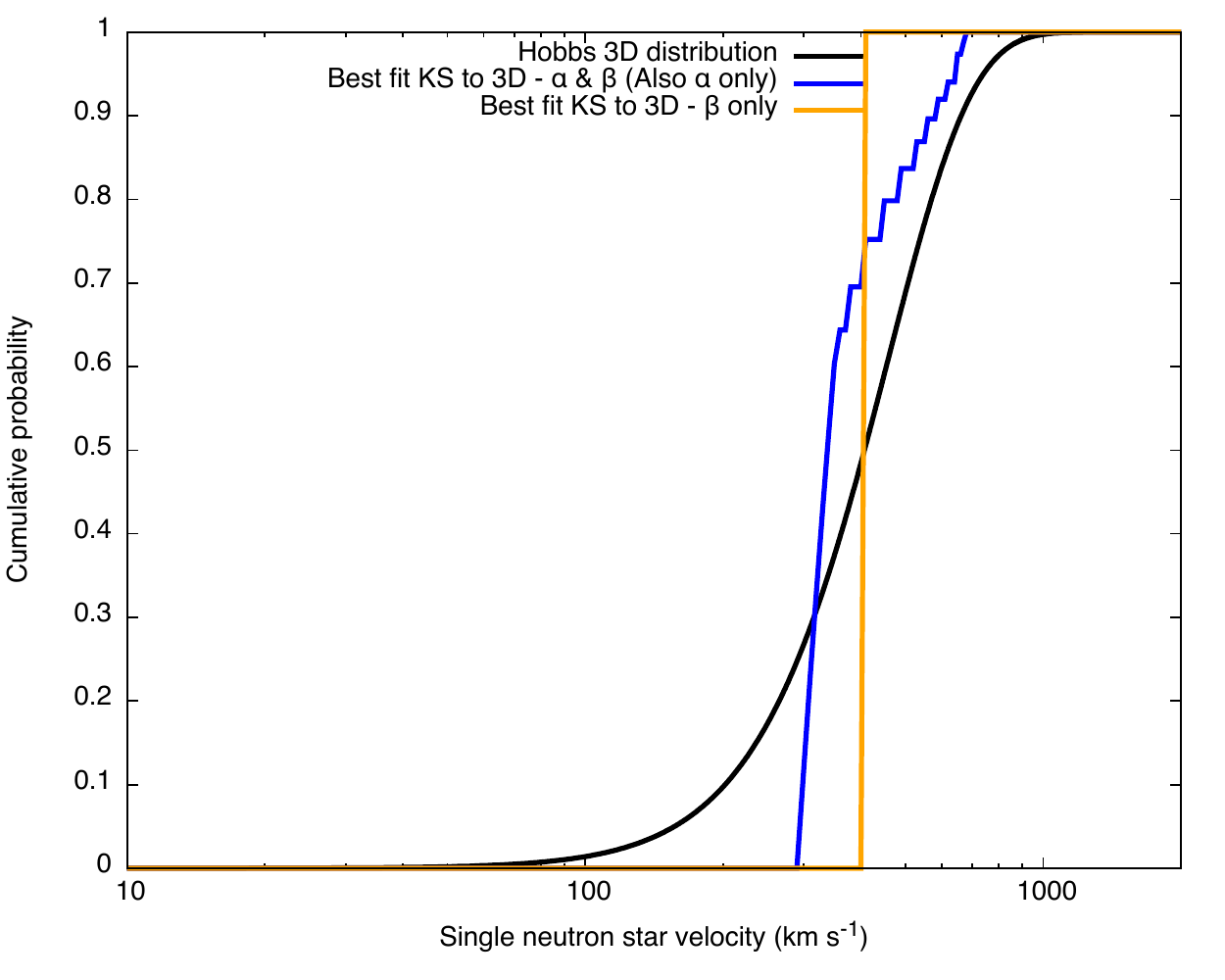}
	\caption{Single stars - `best fit' synthetic three-dimensional (3D) velocity distribution to the Hobbs 3D distribution using the Kolmogorov-Smirnov test - Cumulative probability for $\Gamma$=-2.35, KS `best fit' : $\alpha$ and $\beta$ ($\alpha=70$, $\beta=0$), $\alpha$ ONLY ($\alpha=70$), $\beta$ ONLY, ($\beta$=410)}
	\label{fig:3}
\end{figure}
\begin{figure}
	\vspace{-8mm}
	\hspace{-5mm}
		\includegraphics[scale=0.80]{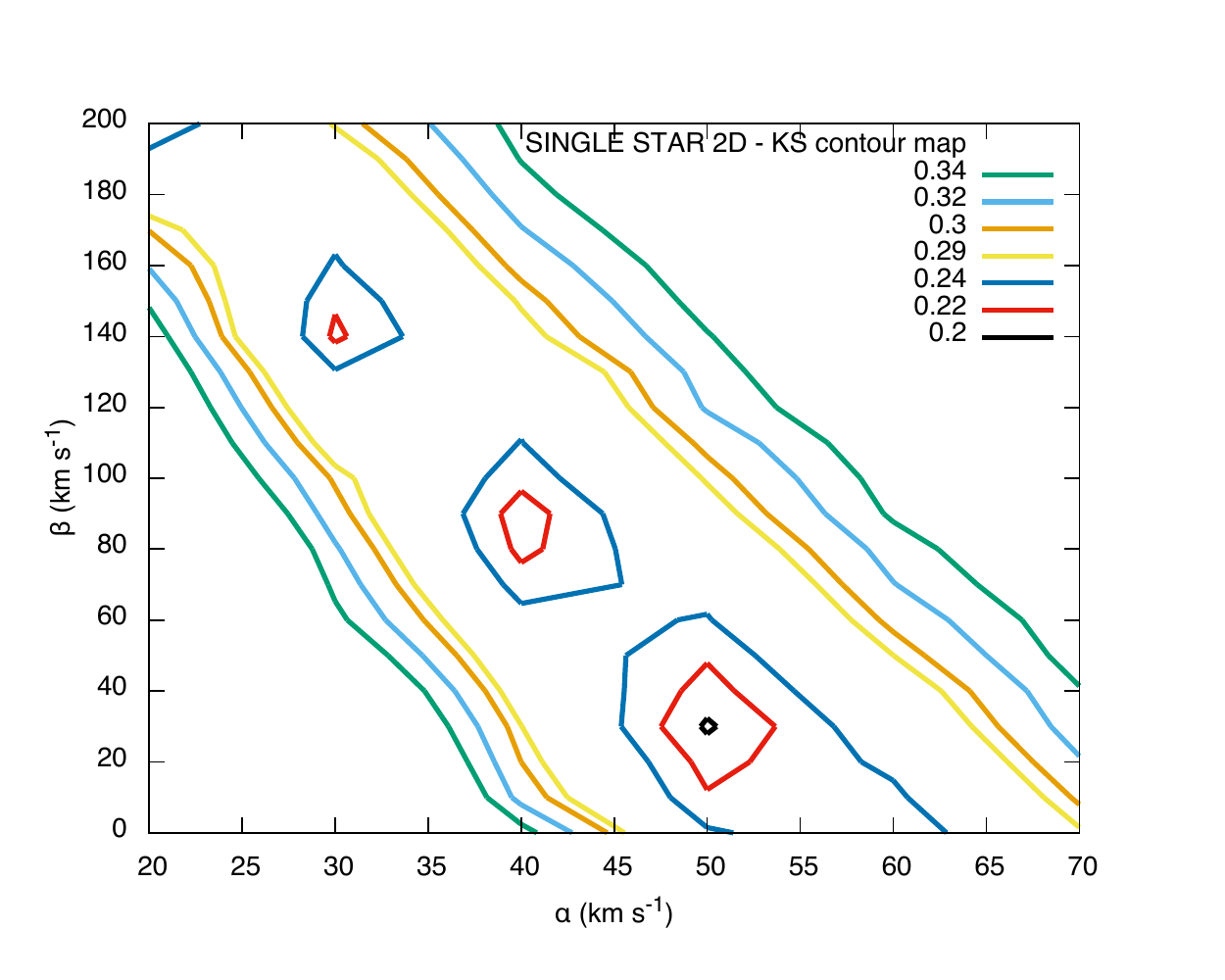}
		\vspace{-4mm}
	\caption{Single stars - Kolmogorov-Smirnov statistic contour plot for synthetic two-dimensional (2D) velocity distribution vs Hobbs 2D subset - $\Gamma$=-2.35, KS `best fit' : $\alpha$=50, $\beta$=30}
	\label{fig:4}
\end{figure}
\section{Results}
\subsection{Single star progenitors}
We show the comparison of our synthetic to observed neutron star velocity distributions for single star progenitors in Figures 2 to 4, with the results of our `best fit' analysis' in Tables 2 and 3. We see that quantitatively, single star progenitors give a poor fit due to the small range of ejecta masses.\\
\\
For the Hobbs 2D subset, the `best fit' is achieved with:
\begin{equation*}
v_{\rm kick} = 50^{+15}_{-10}\,\left(\frac{M_{\rm ejecta}}{M_{\rm remnant}}\right) + 30^{+70}_{-30}
\end{equation*}
This $\alpha$ and $\beta$ combination reaches the critical KS value for 0.05 which is also achieved using a $\beta$ only value of 340 km\,s$^{-1}$. 

Our uncertainties are calculated using the KS statistic contour plot, by recording the values of $\alpha$ and $\beta$ where the $D_{n}$ value enters the 0.01 critical value limit. At this level we can no longer accept the null hypothesis that the two samples are drawn from the same distribution. 

While the `best fit' is achieved with $\alpha=50$ and $\beta=30$ km\,s$^{-1}$, the degeneracy between the variables using single star progenitors is clearly shown in Figure 4, where the $\alpha$ and $\beta$ can be seen to be related. A lower $\alpha$ or $\beta$, is possible but implies that a higher beta or alpha are required respectively.

No $\alpha$ and $\beta$ combination meets the critical value for 0.05 for the Hobbs 3D distribution, with $\alpha=0$ and $\beta=70$ km\,s$^{-1}$ achieving the best result.

Even using our `best fit' values, we find a poor quantitative fit to the Hobbs 2D subset and an extremely poor fit to the Hobbs 3D distribution. While our KS test statistic for the Hobbs 2D subset is less than the 0.05 critical value limit, it fails at the more stringent 0.2 critical value limit, and is poorly defined with significant degeneracy. 

In our single star results, we find a very slight preference for the $\Gamma$ = -2.00 over the $\Gamma$ = -2.35, which is in turn preferred over the $\Gamma$ = -2.70. In all cases the preference is so small that there is no statistical evidence for a preferred IMF.
\begin{table*}	
	\caption{`Best fit' $\alpha$ and $\beta$ variables for three $\Gamma$ values. For each of our isotropic `kick' simulations we present the `best fit' $\alpha$ and $\beta$ values as well as whether the fit achieved the stated Kolmogorov-Smirnov test confidence limit or not. \newline
Key : \ding{51} = meets $D_{n}$ criteria  \ding{55} = fails $D_{n}$ criteria\newline
D$_n$ criteria\newline
Two-dimensional (2D) - $0.05 : \,D_n\leq 0.24$ : $0.20 : \,D_n\leq 0.19$\newline
Three-dimensional (3D) - $0.05 : \,D_n\leq 0.096$ : $0.20 : \,D_n\leq 0.076$}
	\centering	
	\begin{tabular}{l c c c c c c c c c c c r}
		\hline\hline
		IMF ($\Gamma$)  & \multicolumn{4}{l}{-2.00} & \multicolumn{4}{l}{-2.35} & \multicolumn{4}{l}{-2.70}\\
		\multicolumn{3}{l}{KS statistic critical value level $D_{n}$} & $0.05$ & $0.2$ & & & $0.05$ & $0.2$ & & & $0.05$ & $0.2$ \\
		Variable values &  $\alpha\,/{\rm km\,s^{-1}}$ & $\beta\,/{\rm km\,s^{-1}}$  & &  & $\alpha\,/{\rm km\,s^{-1}}$ & $\beta\,/{\rm km\,s^{-1}}$ & & & $\alpha\,/{\rm km\,s^{-1}}$ & $\beta\,/{\rm km\,s^{-1}}$ & \\
		\hline
		\multicolumn{3}{l}{Single star systems} $\alpha$ and $\beta$ - Isotropic `kick'\\
		\hline
		2D points & 50 & 20 & \ding{51} & \ding{55} & 50 & 30 & \ding{51} & \ding{55} & 50 & 40 &  \ding{51} & \ding{55}\\
		3D distribution & 60 & 40 & \ding{55}& \ding{55} & 70 & 0 &  \ding{55} & \ding{55} & 70 & 10 &  \ding{55} & \ding{55}\\
		\hline
		\multicolumn{3}{l}{Binary star systems} $\alpha$ and $\beta$ - Isotropic `kick'\\
		\hline
		2D points & 70 & 110 & \ding{51} & \ding{51} & 70 & 110 & \ding{51} & \ding{51} & 70 & 110 &  \ding{51} & \ding{51}\\
		3D distribution & 70 & 120 & \ding{51}& \ding{51} & 70 & 120 &  \ding{51} & \ding{51} & 70 & 120 &  \ding{51} & \ding{51}\\
		\hline\hline
	\end{tabular}
\end{table*}		
\begin{table*}	
	\caption{`Best fit' $\alpha$ and $\beta$ variables for three $\Gamma$ values. Format is as in Table 2 but with variations of $\alpha$ and $\beta$, models and `kick' direction distributions.\newline	
Key : \ding{51} = meets $D_{n}$ criteria  \ding{55} = fails $D_{n}$ criteria\newline
D$_n$ criteria\newline
Two-dimensional (2D) - $0.05 : \,D_n\leq 0.24$ : $0.20 : \,D_n\leq 0.19$\newline
Three-dimensional (3D) - $0.05 : \,D_n\leq 0.096$ : $0.20 : \,D_n\leq 0.076$}
	\centering	
	\begin{tabular}{l c c c c c c c c c c c r}	
		\hline\hline
		IMF ($\Gamma$)  & \multicolumn{4}{l}{-2.00} & \multicolumn{4}{l}{-2.35} & \multicolumn{4}{l}{-2.70}\\
		\multicolumn{3}{l}{KS statistic critical value level $D_{n}$} & $0.05$ & $0.2$ & & & $0.05$ & $0.2$ & & & $0.05$ & $0.2$ \\
		Variable values &  $\alpha\,/{\rm km\,s^{-1}}$ & $\beta\,/{\rm km\,s^{-1}}$  & &  & $\alpha\,/{\rm km\,s^{-1}}$ & $\beta\,/{\rm km\,s^{-1}}$ & & & $\alpha\,/{\rm km\,s^{-1}}$ & $\beta\,/{\rm km\,s^{-1}}$ & \\
		\hline	
		\multicolumn{4}{l}{Single star $\alpha$ ONLY}\\
		\hline
		2D points & 50 & 0 & \ding{55} & \ding{55} & 60 & 0 & \ding{55} & \ding{55} & 60 & 0 &  \ding{55} & \ding{55}\\
		3D distribution & 70 & 0 & \ding{55}& \ding{55} & 70 & 0 &  \ding{55} & \ding{55} & 70 & 0 &  \ding{55} & \ding{55}\\
		\hline
		\multicolumn{4}{l}{Single star $\beta$ ONLY}\\
		\hline 
		2D points & 0 & 340 & \ding{51} & \ding{55} & 0 & 340 & \ding{51} & \ding{55} & 0 & 340 &  \ding{51} & \ding{55}\\
		3D distribution & 0 & 410 & \ding{55}& \ding{55} & 0 & 410 &  \ding{55} & \ding{55} & 0 & 410 &  \ding{55} & \ding{55}\\
		\hline
		\multicolumn{4}{l}{Binary system $\alpha$ ONLY - Isotropic `kick'}\\
		\hline
		2D points & 90 & 0 & \ding{51} & \ding{51} & 90 & 0 & \ding{51} & \ding{51} & 90 & 0 &  \ding{51} & \ding{51}\\
		3D distribution & 90 & 0 & \ding{51}& \ding{55} & 90 & 0 &  \ding{51} & \ding{55} & 90 & 0 &  \ding{51} & \ding{55}\\
		\hline
		\multicolumn{4}{l}{Binary system $\beta$ ONLY - Isotropic `kick'}\\
		\hline 
		2D points & 0 & 450 & \ding{51} & \ding{51} & 0 & 450 & \ding{51} & \ding{51} & 0 & 450 &  \ding{51} & \ding{51}\\
		3D distribution & 0 & 440 & \ding{55}& \ding{55} & 0 & 440 &  \ding{55} & \ding{55} & 0 & 440 &  \ding{55} & \ding{55}\\
		\hline
		\multicolumn{4}{l}{Binary system $\alpha$ and $\beta$ - Pole `kick'}\\
		\hline 
		2D points & 70 & 120 & \ding{51} & \ding{51} & 70 & 120 & \ding{51} & \ding{51} & 70 & 120 &  \ding{51} & \ding{51}\\
		3D distribution & 60 & 180 & \ding{51}& \ding{51} & 60 & 180 &  \ding{51} & \ding{51} & 70 & 130 &  \ding{51} & \ding{51}\\
		\hline
		\multicolumn{4}{l}{Binary system $\alpha$ and $\beta$ - Equatorial `kick'}\\
		\hline
		2D points & 70  & 100 & \ding{51} & \ding{51} & 70 & 100 & \ding{51} & \ding{51} & 70 & 100 &  \ding{51} & \ding{51}\\
		3D distribution & 70 & 110 & \ding{51}& \ding{51} & 70 & 110 &  \ding{51} & \ding{51} & 60 & 110 &  \ding{51} & \ding{51}\\
		\hline\hline
		
	\end{tabular}
\end{table*}
\begin{figure}
		\includegraphics[scale=0.72]{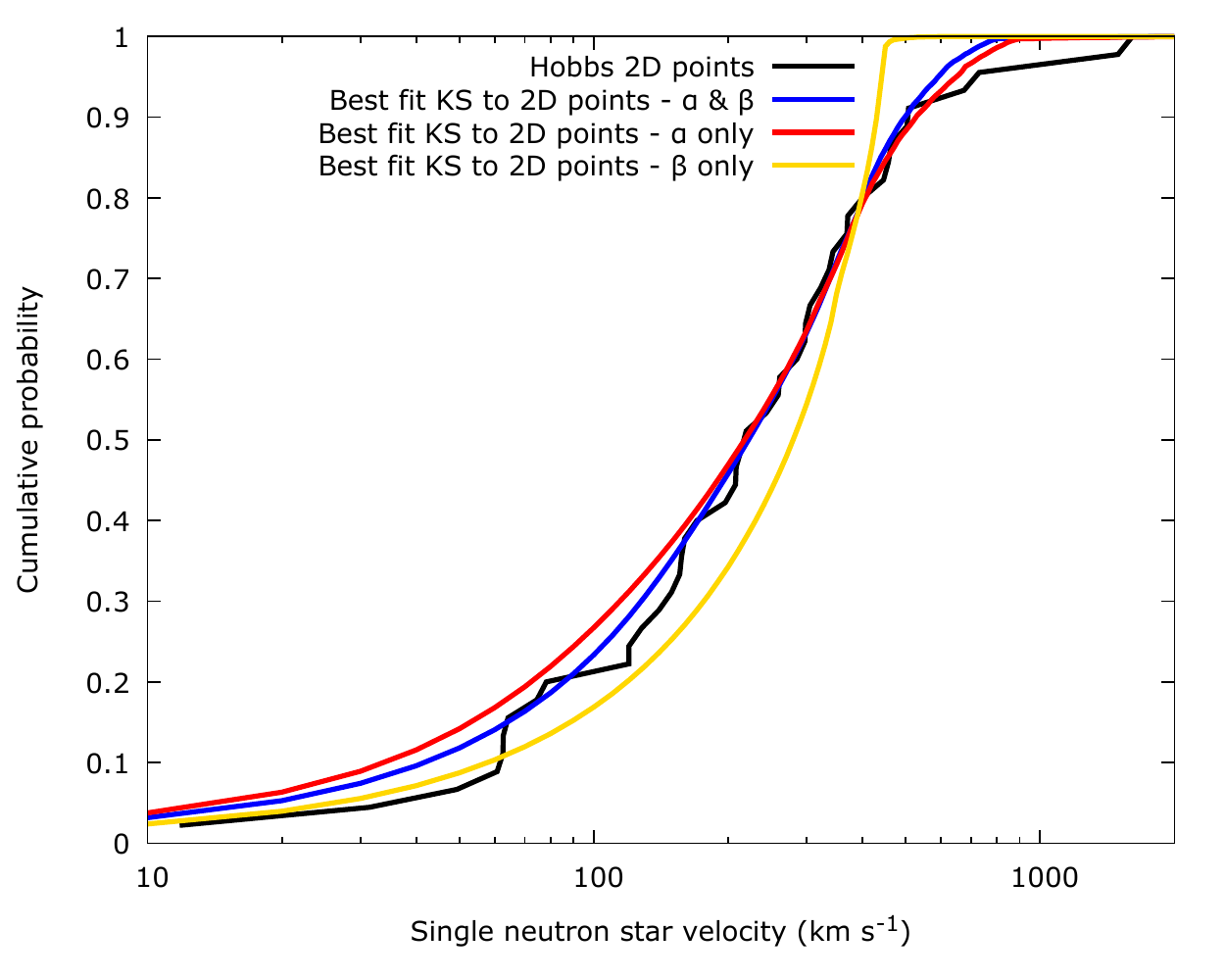}
		\vspace{0mm}
	\caption{Binary stars - `Best fit' synthetic two-dimensional (2D) velocity distribution to the Hobbs 2D subset using the Kolmogorov-Smirnov test - Cumulative probability for $\Gamma$=-2.35, KS `best fit' : $\alpha$ and $\beta$ ($\alpha$=70, $\beta$=110), $\alpha$ ONLY ($\alpha$=90), $\beta$ ONLY ($\beta$=450)}
	\label{fig:5}
\end{figure}
\begin{figure}
	\vspace{-8mm}
	\hspace{-5mm}
		\includegraphics[scale=0.80]{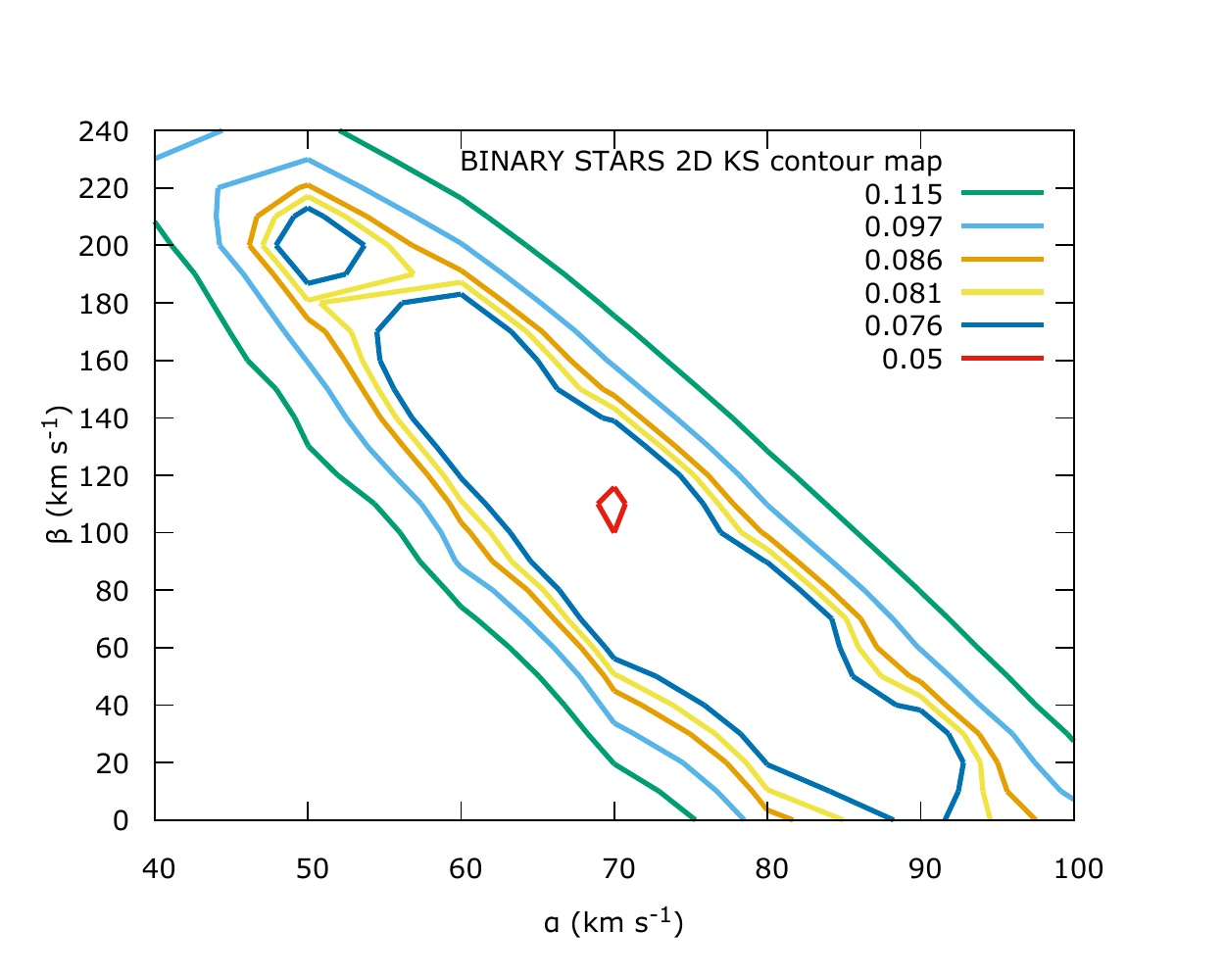}
		\vspace{-6mm}
	\caption{Binary stars - Kolmogorov-Smirnov statistic contour plot for synthetic two-dimensional (2D) velocity distribution vs Hobbs 2D subset - $\Gamma$=-2.35, KS `best fit' : $\alpha$=70, $\beta$=110}
	\label{fig:6}
\end{figure}
\begin{figure}
	\vspace{-0mm}
	\hspace{-2mm}
		\includegraphics[scale=0.72]{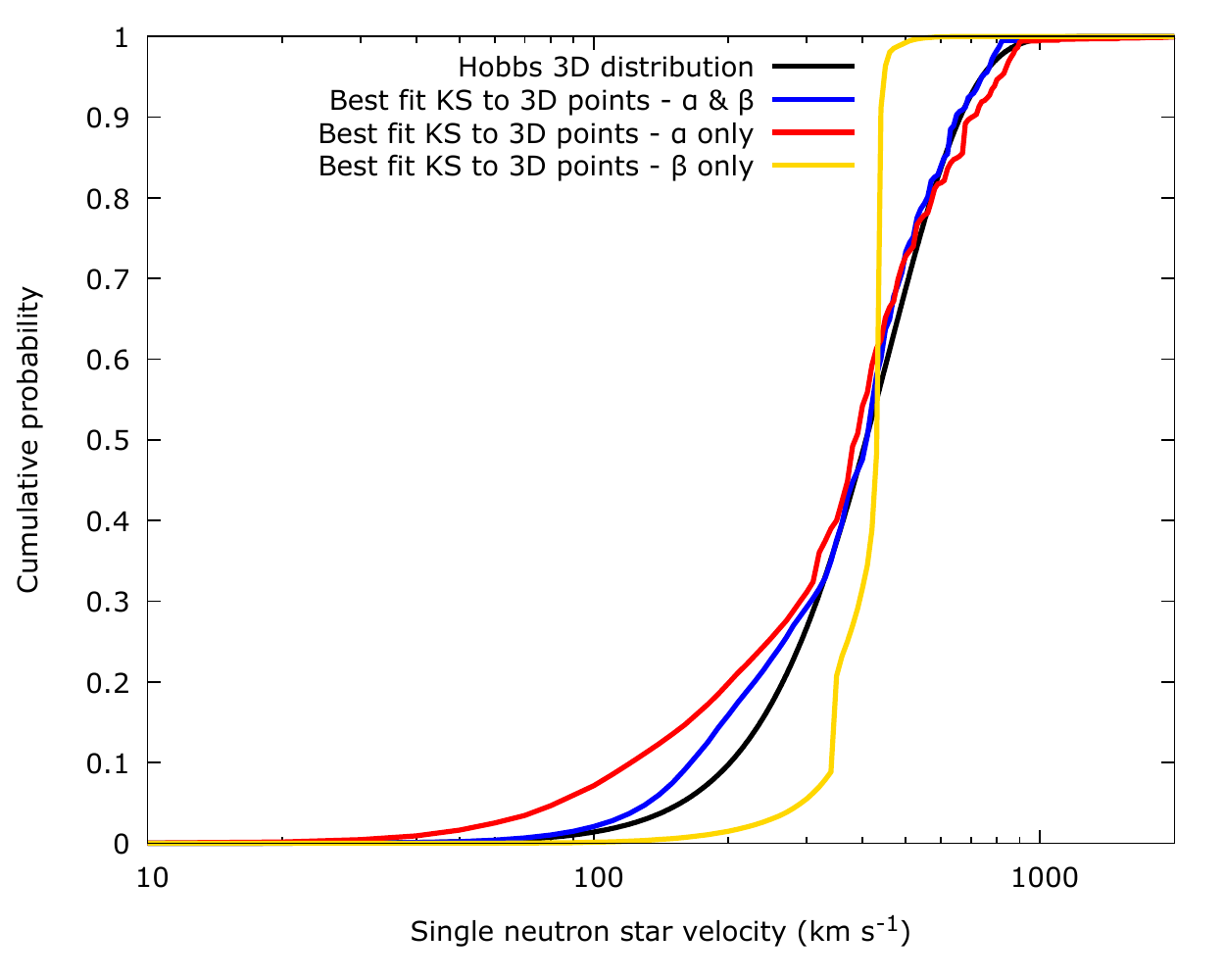}
	\caption{Binary stars - `Best fit' synthetic three-dimensional (3D) velocity distribution to the Hobbs 3D distribution using the Kolmogorov-Smirnov test - Cumulative probability for $\Gamma$=-2.35, KS `best fit' : $\alpha$ and $\beta$ ($\alpha$=70, $\beta$=120), $\alpha$ ONLY ($\alpha$=90), $\beta$ ONLY ($\beta$=440)}
	\label{fig:7}
\end{figure}
\begin{figure}
	\vspace{-8mm}
	\hspace{-5mm}
		\includegraphics[scale=0.80]{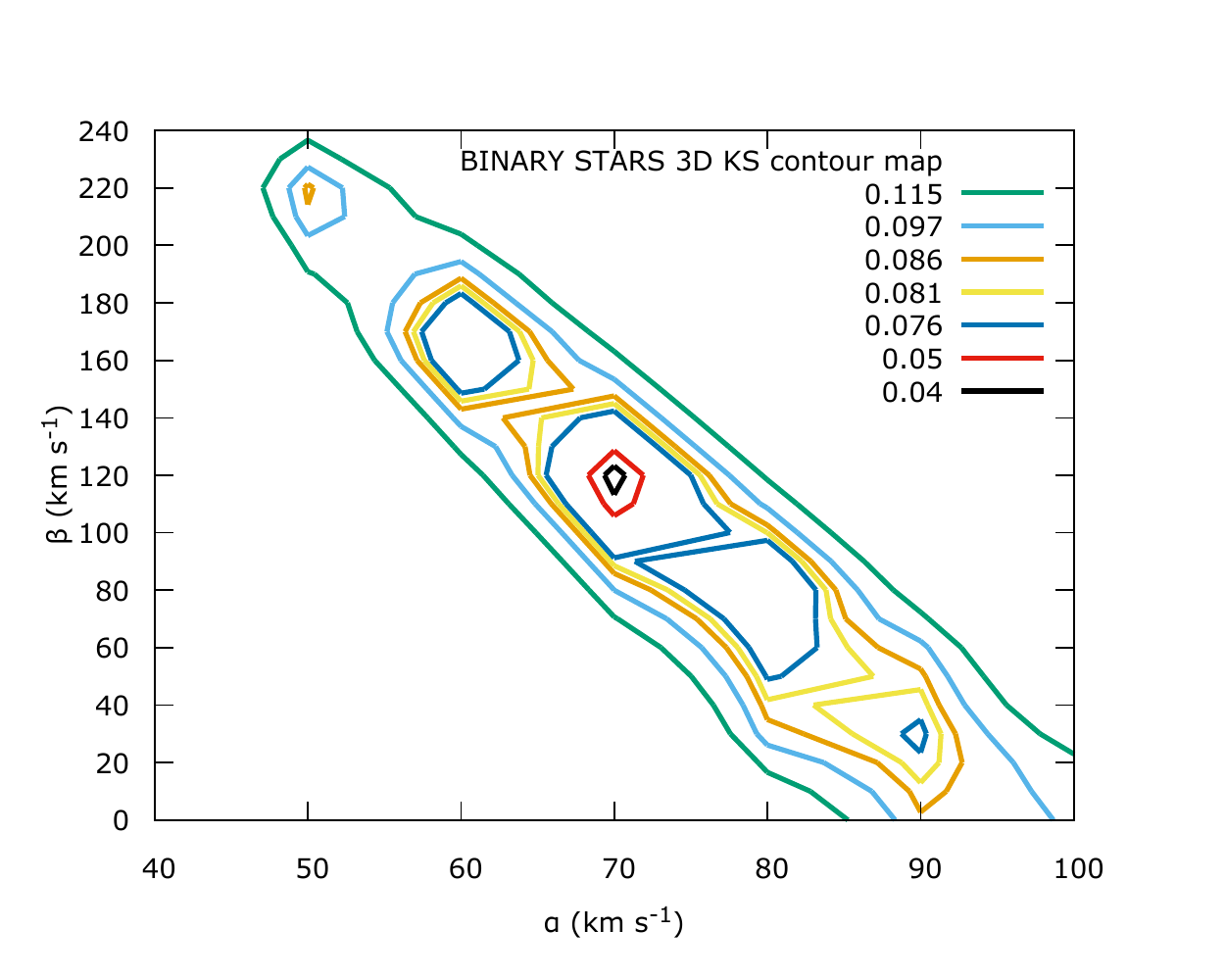}
		\vspace{-5mm}
	\caption{Binary stars - Kolmogorov-Smirnov statistic contour plot for synthetic three-dimensional (3D) velocity distribution vs Hobbs 3D distribution - $\Gamma$=-2.35, KS `best fit' : $\alpha$=70, $\beta$=120}
	\label{fig:8}
\end{figure}

\subsection{Binary star progenitors}
We show the comparison of our synthetic to observed neutron star velocity distributions for binary star progenitors in Figures 5 to 8, with the results of our `best fit' analysis' in Tables 2 and 3. In contrast to the single star progenitors, we see an excellent qualitative fit, due to the binary interactions providing a wider range of ejecta masses.\\
\\
For the Hobbs 2D subset the `best fit' is achieved with:
\begin{equation*}
v_{\rm kick} = 70^{+15}_{-15}\,\left(\frac{M_{\rm ejecta}}{M_{\rm remnant}}\right) + 110^{+70}_{-90}
\end{equation*}
Our `best fit' to the Hobbs 3D distribution is achieved with:
\begin{equation*}
v_{\rm kick} = 70^{+10}_{-10}\,\left(\frac{M_{\rm ejecta}}{M_{\rm remnant}}\right) + 120^{+45}_{-50}
\end{equation*}

As in the single star progenitors, the uncertainties on the $\alpha$ and $\beta$ variables are calculated using the KS statistic contour plot.

Our `best fit' combinations provide an excellent fit to both the Hobbs 2D subset and the Hobbs 3D distribution, with the general shape and range well reproduced. 

The `best fit' using $\alpha$ only also meets both the KS critical value for 0.05 and 0.2 for the Hobbs 2D subset, for all of the IMFs tested, and in addition, reaches the critical value for 0.05 for the Hobbs 3D distribution.
 
The polar and equatorial `kick' orientations provide an almost identical fit with very similar variables and again the KS statistic meets both the 0.05 and 0.2 critical value limits. While the KS statistic was lower for the isotropic `kick', when compared to both the polar and equatorial `kick' orientations, once again the difference is so small that there is no statistical evidence for a preferred `kick' orientation.

We find the KS statistics show a very slight preference for the $\Gamma$ = -2.00 over the $\Gamma$ = -2.35, which is in turn preferred over the $\Gamma$ = -2.70. Again the preference is so small that it provides no statistical evidence for a preferred IMF.

\section{Discussion}
Given that between 50 and 80 per cent of massive stars are thought to be in binaries \citep{RN277,RN304,RN205}, it is of little surprise that our synthetic velocity distributions generated by supernovae in single star progenitors shows a very poor fit to the observational data. What is somewhat surprising, is just how well our simple `kick' model, using the binary star progenitors, fits the data.

Using our synthetic binary star population, our kick model and an isotropic `kick' orientation, our synthetic 2D velocity distribution is an excellent qualitative fit to the Hobbs 2D subset, with the possible exception of the high velocity points (over 600 km\,s$^{-1}$). Our 3D synthetic distribution is also an excellent fit, both qualitatively and quantitatively, to the Hobbs 3D distribution, throughout the entire velocity range. 

For binaries, the `best fit' variables for both the synthetic 2D and 3D distributions also surpass the more stringent KS 0.2 critical value limit. The fact that for our binary systems, we obtain almost identical `best fit' variables for both the Hobbs 2D subset and Hobbs 3D distribution, is reassuring and confirms that our Hobbs 2D subset is not significantly different to the points used by \cite{RN165} to derive their 3D distribution.

The poorer fit of our synthetic distributions to the Hobbs 2D subset, while our 3D distributions still fit the Hobbs 3D distribution, would seem to support the claim by \cite{RN224} that the `deconvolution technique' used by \cite{RN165} lost some information particularly in the high-velocity tail of the distribution. We also note that the \cite{RN165} dataset, and consequently, our own subset, may have a possible bias against low velocity neutron stars, since these low proper motions would be difficult to detect.

The fact that there is little difference between the `best fit' $\alpha$ and $\beta$ values derived from our different `kick' orientations is interesting. It indicates that the most important parameter in determining the neutron star velocity distribution is the distribution of ejecta masses in the stellar models. The reason that single star progenitors provide a poor fit, is because they have only a restricted range of ejecta masses, with most progenitors being red supergiants with massive hydrogen envelopes. In contrast, the binary populations have a range of ejecta masses from less than 1 $M_{\odot}$, up to more than 10 $M_{\odot}$ for wider binaries that effectively evolve as single stars. The degeneracy between $\alpha$ and $\beta$ values in our `best fit' analysis' also indicates that even if our `best fit' values are modified, and should be picked from some random distribution, it is still the ejecta mass distribution that determines the neutron star `kick' velocity distribution.

The presence of a significant `kick', in the same direction as the conservation of momentum `kick', is an interesting feature. One possible explanation could be that a small net anisotropy in the neutrino emission or neutrino flavour emission, as proposed by \cite{RN294}, creates the 110 / 120 kms$^{-1}$ $\beta$ recoil velocity. This is of the order of the `several 10 km s$^{-1}$' they suggest. The subsequent creation of a region behind the stalled shock with higher neutrino density, could lead to the large-scale asymmetry in the ejecta, as suggested by \cite{RN222}. The resulting conservation of momentum then explaining our $\alpha$ velocity component of 70 km\,s$^{-1}$. 

\section{Conclusions}
We proposed that the velocity of neutron stars is determined, primarily, by the mass ejected from the supernovae that created them. 

We used BPASS models, and our REAPER code to create velocity distributions, for a range of $\alpha$ and $\beta$ values, for single neutron stars created in supernovae, from both single and binary star progenitors. We examined three different IMFs and three different `kick' orientations. We compared our synthetic 2D and 3D distributions to our Hobbs 2D subset and the Hobbs 3D distribution, using the KS statistic, and found the `best fit' to the 2D data (for $\Gamma$= -2.35) to be from binary star system progenitors using: 

\begin{equation*}
v_{\rm kick} = 70^{+15}_{-15}\,\left(\frac{M_{\rm ejecta}}{M_{\rm remnant}}\right) + 110^{+70}_{-90}
\end{equation*}

While the KS statistic showed the isotropic `kick' was a better fit to the observational data than the pole or equatorial `kicks' suggested by \cite{RN203}, the difference was so small that it provides no statistical evidence of a preferred `kick' orientation.

The most shallow IMF ($\Gamma$=-2.00) was shown by the KS statistic to be the `best fit', but again the difference was so small that it provides no statistical evidence of a preferred IMF.

There are a number of mature binary population synthesis codes in use around the world, which could utilise a `kick' model based on the physical attributes of the progenitors \citep[e.g.][]{RN147,RN256,RN255,RN257,RN258}. Typically, the neutron star `kick' velocity is chosen at random from a distribution, but here we suggest this physically motivated `kick' velocity should be adopted as more accurately reflecting what occurs in nature.

The `best fit' (before penalising complexity), to the Hobbs 2D subset was achieved with a bi-modal distribution. Binary systems naturally produce bi-modal velocity distributions with supernovae in merged and `runaway' stars generally producing compact remnants with higher velocities, while those produced in the disruption of the binary producing compact remnants of lower velocities.

The `best fits' for both the Hobbs 2D subset and the Hobbs 3D distribution, were achieved with almost identical $\alpha$ and $\beta$ variables, implying consistency in both dimensions.

Our `kick' formula provides a direct link between the compact remnant velocity and the supernova ejecta mass and hence provides more detailed velocity information on these objects and the binaries that contain them. 

The key benefit we see to utilising our proposed relationship, is that it gives velocity information on individual neutron stars, based on the physical properties of their progenitors, rather than on a random basis.

In summary, our simulations suggest that neutron star `kicks' (and hence their final velocities), are likely to be linked to, but not solely determined by, the amount of material ejected in their natal supernovae. If further work indicates this to be true then it is likely that high velocity neutron stars are created in wide binaries, while lower velocity neutron stars, such as those in surviving binary systems, arise from closer binaries where the ejecta mass is lost in a binary interaction with the companion. Also that the `kick' velocity alone may give an indication of the progenitor mass for the neutron star.

While we have examined one observable test, the model needs to undergo further observational tests to provide stronger evidence of its validity. 

\section{Acknowledgements}
We would like to thank the anonymous referee for his insightful suggestions that enabled us to improve this paper.

JCB acknowledges support from the University of Auckland and in particular Dr. Brendon Brewer for his assistance in deriving the Maxwell-Boltzmann scaling factor.

The authors wish to acknowledge the contribution of the NeSI high-performance computing facilities and the staff at the Centre for eResearch at the University of Auckland. New Zealand's national facilities are provided by the New Zealand eScience Infrastructure (NeSI) and funded jointly by NeSI's collaborator institutions and through the Ministry of Business, Innovation and Employment's Infrastructure programme. URL: http://www.nesi.org.nz

\bsp
\addtocontents{toc}{\protect\vspace*{\baselineskip}}
\bibliography{./KickPaperV2.bib}

\begin{thebibliography}{}

\bibitem[\protect\citeauthoryear{Arzoumanian, Chernoff \& Cordes}{Arzoumanian
  et~al.}{2002}]{RN215}
Arzoumanian Z.,  Chernoff D.~F.,    Cordes J.~M.,  2002, ApJ, 568, 289

\bibitem[\protect\citeauthoryear{Budiardja, Cardall \& Endeve}{Budiardja
  et~al.}{2015}]{RN307}
Budiardja R.,  Cardall C.,    Endeve E.,  2015, ArXiv e-prints

\bibitem[\protect\citeauthoryear{Chini, Hoffmeister, Nasseri, Stahl \&
  Zinnecker}{Chini et~al.}{2012}]{RN304}
Chini R.,  Hoffmeister V.,  Nasseri A.,  Stahl O.,    Zinnecker H.,  2012,
  MNRAS, 424, 1925

\bibitem[\protect\citeauthoryear{Couch \& Ott}{Couch \& Ott}{2013}]{RN308}
Couch S.,  Ott C.,  2013, ApJ, 778, L7

\bibitem[\protect\citeauthoryear{Delgado-Donate, Clarke, Bate \&
  Hodgkin}{Delgado-Donate et~al.}{2004}]{RN277}
Delgado-Donate E.~J.,  Clarke C.~J.,  Bate M.~R.,    Hodgkin S.~T.,  2004,
  MNRAS, 351, 617

\bibitem[\protect\citeauthoryear{Eggleton}{Eggleton}{1971}]{RN199}
Eggleton P.~P.,  1971, MNRAS, 151, 351

\bibitem[\protect\citeauthoryear{Eldridge \& Tout}{Eldridge \&
  Tout}{2005}]{RN286}
Eldridge J.,  Tout C.,  2005, in Turatto M.,  Benetti S.,  Zampieri L.,   Shea
  W.,  eds, 1604-2004: Supernovae as Cosmological Lighthouses Vol.~342 of
  Astronomical Society of the Pacific Conference Series, Modelling the
  progenitors of core-collapse supernovae.
p.~126

\bibitem[\protect\citeauthoryear{Eldridge, Izzard \& Tout}{Eldridge
  et~al.}{2008}]{RN270}
Eldridge J.~J.,  Izzard R.~G.,    Tout C.~A.,  2008, MNRAS, 384, 1109

\bibitem[\protect\citeauthoryear{Eldridge, Langer \& Tout}{Eldridge
  et~al.}{2011}]{RN207}
Eldridge J.~J.,  Langer N.,    Tout C.~A.,  2011, MNRAS, 414, 3501

\bibitem[\protect\citeauthoryear{Faucher-Giguere \& Kaspi}{Faucher-Giguere \&
  Kaspi}{2006}]{RN224}
Faucher-Giguere C.~A.,  Kaspi V.~M.,  2006, ApJ, 643, 332

\bibitem[\protect\citeauthoryear{Fryer, Burrows \& Benz}{Fryer
  et~al.}{1998}]{RN126}
Fryer C.,  Burrows A.,    Benz W.,  1998, ApJ, 496, 333

\bibitem[\protect\citeauthoryear{Gvaramadze}{Gvaramadze}{2007}]{RN305}
Gvaramadze V.,  2007, A\&A, 470, L9

\bibitem[\protect\citeauthoryear{Hansen \& Phinney}{Hansen \&
  Phinney}{1997}]{RN213}
Hansen B. M.~S.,  Phinney E.~S.,  1997, MNRAS, 291, 569

\bibitem[\protect\citeauthoryear{Hix, Lentz, Bruenn, Mezzacappa, Messer,
  Endeve, Blondin, Harris, Marronetti \& Yakunin}{Hix et~al.}{2016}]{RN306}
Hix W.,  Lentz E.,  Bruenn S.~W.,  Mezzacappa A.,  Messer O.,  Endeve E.,
  Blondin J.,  Harris J.,  Marronetti P.,    Yakunin K.,  2016, Acta Physica
  Polonica B, 47, 645

\bibitem[\protect\citeauthoryear{Hobbs, Lorimer, Lyne \& Kramer}{Hobbs
  et~al.}{2005}]{RN165}
Hobbs G.,  Lorimer D.~R.,  Lyne A.~G.,    Kramer M.,  2005, MNRAS, 360, 974

\bibitem[\protect\citeauthoryear{Hurley, Tout \& Pols}{Hurley
  et~al.}{2002}]{RN255}
Hurley J.,  Tout C.,    Pols O.,  2002, MNRAS, 329, 897

\bibitem[\protect\citeauthoryear{Izzard, Ramirez-Ruiz \& Tout}{Izzard
  et~al.}{2004}]{RN257}
Izzard R.~G.,  Ramirez-Ruiz E.,    Tout C.~A.,  2004, MNRAS, 348, 1215

\bibitem[\protect\citeauthoryear{Janka \& Muller}{Janka \&
  Muller}{1994}]{RN238}
Janka H.~T.,  Muller E.,  1994, A\&A, 290, 496

\bibitem[\protect\citeauthoryear{Kiminki \& Kobulnicky}{Kiminki \&
  Kobulnicky}{2012}]{RN264}
Kiminki D.~C.,  Kobulnicky H.~A.,  2012, ApJ, 751

\bibitem[\protect\citeauthoryear{Knigge, Coe \& Podsiadlowski}{Knigge
  et~al.}{2011}]{RN278}
Knigge C.,  Coe M.~J.,    Podsiadlowski P.,  2011, Nature, 479, 372

\bibitem[\protect\citeauthoryear{Kroupa, Tout \& Gilmore}{Kroupa
  et~al.}{1993}]{RN179}
Kroupa P.,  Tout C.~A.,    Gilmore G.,  1993, MNRAS, 262, 545

\bibitem[\protect\citeauthoryear{Lattimer \& Prakash}{Lattimer \&
  Prakash}{2005}]{RN246}
Lattimer J.~M.,  Prakash M.,  2005, Phys.Rev.Lett., 94, 111101

\bibitem[\protect\citeauthoryear{Lipunov \& Pruzhinskaya}{Lipunov \&
  Pruzhinskaya}{2014}]{RN258}
Lipunov V.~M.,  Pruzhinskaya M.~V.,  2014, MNRAS, 440, 1193

\bibitem[\protect\citeauthoryear{Liu, Tauris, R{\"o}pke, Moriya, Kruckow,
  Stancliffe \& Izzard}{Liu et~al.}{2015}]{RN186}
Liu Z.,  Tauris T.,  R{\"o}pke F.,  Moriya T.,  Kruckow M.,  Stancliffe R.,
  Izzard R.,  2015, A\&A, 584, A11

\bibitem[\protect\citeauthoryear{Lyne \& Lorimer}{Lyne \&
  Lorimer}{1994}]{RN155}
Lyne A.~G.,  Lorimer D.~R.,  1994, Nature, 369, 127

\bibitem[\protect\citeauthoryear{Ma, Hopkins, Faucher-Giguere, Zolman, Muratov,
  Keres \& Quataert}{Ma et~al.}{2016}]{RN283}
Ma X.,  Hopkins P.~F.,  Faucher-Giguere C.-A.,  Zolman N.,  Muratov A.~L.,
  Keres D.,    Quataert E.,  2016, MNRAS, 456, 2140

\bibitem[\protect\citeauthoryear{Mennekens \& Vanbeveren}{Mennekens \&
  Vanbeveren}{2014}]{RN287}
Mennekens N.,  Vanbeveren D.,  2014, A\&A, 564, A134

\bibitem[\protect\citeauthoryear{Moriya \& Eldridge}{Moriya \&
  Eldridge}{2016}]{RN290}
Moriya T.~J.,  Eldridge J.~J.,  2016, ArXiv e-prints

\bibitem[\protect\citeauthoryear{Nieva \& Przybilla}{Nieva \&
  Przybilla}{2012}]{RN289}
Nieva M.-F.,  Przybilla N.,  2012, A\&A, 539, A143

\bibitem[\protect\citeauthoryear{Noutsos, Schnitzeler, Keane, Kramer \&
  Johnston}{Noutsos et~al.}{2013}]{RN203}
Noutsos A.,  Schnitzeler D.,  Keane E.,  Kramer M.,    Johnston S.,  2013,
  MNRAS, 430, 2281

\bibitem[\protect\citeauthoryear{Orlando, Miceli, Pumo \& Bocchino}{Orlando
  et~al.}{2016}]{RN251}
Orlando S.,  Miceli M.,  Pumo M.,    Bocchino F.,  2016, ArXiv e-prints

\bibitem[\protect\citeauthoryear{Pfahl, Rappaport, Podsiadlowski \&
  Spruit}{Pfahl et~al.}{2002}]{RN239}
Pfahl E.,  Rappaport S.,  Podsiadlowski P.,    Spruit H.,  2002, ApJ, 574, 364

\bibitem[\protect\citeauthoryear{Podsiadlowski, Langer, Poelarends, Rappaport,
  Heger \& Pfahl}{Podsiadlowski et~al.}{2004}]{RN144}
Podsiadlowski P.,  Langer N.,  Poelarends A. J.~T.,  Rappaport S.,  Heger A.,
   Pfahl E.,  2004, ApJ, 612, 1044

\bibitem[\protect\citeauthoryear{Prieto, Lambert \& Asplund}{Prieto
  et~al.}{2002}]{RN288}
Prieto C.~A.,  Lambert D.~L.,    Asplund M.,  2002, ApJ, 573, L137

\bibitem[\protect\citeauthoryear{Rest, Foley, Sinnott, Welch, Badenes,
  Filippenko, Bergmann, Bhatti, Blondin, Challis P., Damke, Finley, Huber,
  Kasen, Kirshner, Matheson, Mazzali, Minniti, Nakajima, Narayan, Olsen, Sauer,
  Smith \& Suntzeff}{Rest et~al.}{2011}]{RN242}
Rest A.,  Foley R.,  Sinnott B.,  Welch D.,  Badenes C.,  Filippenko A.,
  Bergmann M.,  Bhatti W.,  Blondin S.,  Challis P. P.,  Damke G.,  Finley H.,
  Huber M.,  Kasen D.,  Kirshner R.,  Matheson T.,  Mazzali P.,  Minniti D.,
  Nakajima R.,  Narayan G.,  Olsen K.,  Sauer D.,  Smith R.,    Suntzeff N.,
  2011, ApJ, 732, 3

\bibitem[\protect\citeauthoryear{Sana, de Koter, de Mink, Dunstall, Evans,
  Henault-Brunet, Maiz~Apellaniz, Ramirez-Agudelo, Taylor, Walborn, Clark,
  Crowther, Herrero, Gieles, Langer, Lennon \& Vink}{Sana et~al.}{2013}]{RN205}
Sana H.,  de Koter A.,  de Mink S.,  Dunstall P.,  Evans C.,  Henault-Brunet
  V.,  Maiz~Apellaniz J.,  Ramirez-Agudelo O.,  Taylor W.,  Walborn N.,  Clark
  J.,  Crowther P.,  Herrero A.,  Gieles M.,  Langer N.,  Lennon D.,    Vink
  J.,  2013, AAP, 550, A107

\bibitem[\protect\citeauthoryear{Sana, de Mink, de Koter, Langer, Evans,
  Gieles, Gosset, Izzard, Le~Bouquin \& Schneider}{Sana et~al.}{2013}]{RN206}
Sana H.,  de Mink S.,  de Koter A.,  Langer N.,  Evans C.,  Gieles M.,  Gosset
  E.,  Izzard R.,  Le~Bouquin J.,    Schneider F.,  2013, in Pugliese G.,  de
  Koter A.,   Wijburg M.,  eds, 370 Years of Astronomy in Utrecht Vol.~470 of
  Astronomical Society of the Pacific Conference Series, Multiplicity of
  massive o stars and evolutionary implications.
p.~141

\bibitem[\protect\citeauthoryear{Scalo}{Scalo}{1986}]{RN284}
Scalo J.~M.,  1986, FCP, 11, 1

\bibitem[\protect\citeauthoryear{Scheck, Kifonidis, Janka \& Mueller}{Scheck
  et~al.}{2006}]{RN235}
Scheck L.,  Kifonidis K.,  Janka H.~T.,    Mueller E.,  2006, A\&A, 457, 963

\bibitem[\protect\citeauthoryear{Smartt}{Smartt}{2009}]{RN202}
Smartt S.,  2009, ARA\&A, 47, 63

\bibitem[\protect\citeauthoryear{Smartt}{Smartt}{2015}]{RN140}
Smartt S.,  2015, PASA, 32, e016

\bibitem[\protect\citeauthoryear{Stanway, Eldridge \& Becker}{Stanway
  et~al.}{2016}]{RN280}
Stanway E.~R.,  Eldridge J.~J.,    Becker G.~D.,  2016, MNRAS

\bibitem[\protect\citeauthoryear{Tamborra, Hanke, Janka, M{\"u}ller, Raffelt \&
  Marek}{Tamborra et~al.}{2014}]{RN294}
Tamborra I.,  Hanke F.,  Janka H.,  M{\"u}ller B.,  Raffelt G.,    Marek A.,
  2014, ApJ, 792, 96

\bibitem[\protect\citeauthoryear{Tauris, Langer, Moriya, Podsiadlowski, Yoon \&
  Blinnikov}{Tauris et~al.}{2013}]{RN309}
Tauris T.,  Langer N.,  Moriya T.,  Podsiadlowski P.,  Yoon S.,    Blinnikov
  S.,  2013, ApJ, 778, L23

\bibitem[\protect\citeauthoryear{Tauris, Langer \& Podsiadlowski}{Tauris
  et~al.}{2015}]{RN310}
Tauris T.,  Langer N.,    Podsiadlowski P.,  2015, MNRAS, 451, 2123

\bibitem[\protect\citeauthoryear{Tauris, Sanyal, Yoon \& Langer}{Tauris
  et~al.}{2013}]{RN204}
Tauris T.,  Sanyal D.,  Yoon S.,    Langer N.,  2013, A\&A, 558, A39

\bibitem[\protect\citeauthoryear{Tauris \& Bailes}{Tauris \&
  Bailes}{1996}]{RN147}
Tauris T.~M.,  Bailes M.,  1996, A\&A, 315, 432

\bibitem[\protect\citeauthoryear{Tauris, Fender, van~den Heuvel, Johnston \&
  Wu}{Tauris et~al.}{1999}]{RN194}
Tauris T.~M.,  Fender R.~P.,  van~den Heuvel E. P.~J.,  Johnston H.~M.,    Wu
  K.,  1999, MNRAS, 310, 1165

\bibitem[\protect\citeauthoryear{Tauris \& Takens}{Tauris \&
  Takens}{1998}]{RN146}
Tauris T.~M.,  Takens R.~J.,  1998, A\&A, 330, 1047

\bibitem[\protect\citeauthoryear{Vagnozzi, Freese \& Zurbuchen}{Vagnozzi
  et~al.}{2016}]{RN252}
Vagnozzi S.,  Freese K.,    Zurbuchen T.~H.,  2016, ArXiv e-prints

\bibitem[\protect\citeauthoryear{Vanbeveren, De~Donder, Van~Bever,
  Van~Rensbergen \& De~Loore}{Vanbeveren et~al.}{1998}]{RN285}
Vanbeveren D.,  De~Donder E.,  Van~Bever J.,  Van~Rensbergen W.,    De~Loore
  C.,  1998, NA, 3, 443

\bibitem[\protect\citeauthoryear{Vanbeveren, De~Loore \&
  Van~Rensbergen}{Vanbeveren et~al.}{1998}]{RN256}
Vanbeveren D.,  De~Loore C.,    Van~Rensbergen W.,  1998, A\&A Rev., 9, 63

\bibitem[\protect\citeauthoryear{Villante, Serenelli, Delahaye \&
  Pinsonneault}{Villante et~al.}{2013}]{RN279}
Villante F.,  Serenelli A.,  Delahaye F.,    Pinsonneault M.,  2013, ApJ, 787,
  13

\bibitem[\protect\citeauthoryear{Wenger, Ochsenbein, Egret, Dubois, Bonnarel,
  Borde, Genova, Jasniewicz, Laloe, Lesteven \& Monier}{Wenger
  et~al.}{2000}]{RN268}
Wenger M.,  Ochsenbein F.,  Egret D.,  Dubois P.,  Bonnarel F.,  Borde S.,
  Genova F.,  Jasniewicz G.,  Laloe S.,  Lesteven S.,    Monier R.,  2000,
  AAPS, 143, 9

\bibitem[\protect\citeauthoryear{Wilkins, Feng, Matteo, Croft, Stanway, Bouwens
  \& Thomas}{Wilkins et~al.}{2016}]{RN282}
Wilkins S.~M.,  Feng Y.,  Matteo T.~D.,  Croft R.,  Stanway E.~R.,  Bouwens
  R.~J.,    Thomas P.,  2016, MNRAS, 458, L6

\bibitem[\protect\citeauthoryear{Wofford \& et al}{Wofford \&
  et~al}{2016}]{RN281}
Wofford A.,  et al 2016, MNRAS, 457, 4296

\bibitem[\protect\citeauthoryear{Wongwathanarat, Janka \&
  Muller}{Wongwathanarat et~al.}{2010}]{RN222}
Wongwathanarat A.,  Janka H.~T.,    Muller E.,  2010, ApJ, 725, L106

\end{thebibliography}

\begin{appendices}
\renewcommand\thesection{\arabic{section}}
\section{Derivation of modified Maxwell-Boltzmann scaling factor}
We know that the uni-variate Maxwell-Boltzmann tangential velocity distribution ($v_{xy}$) is comprised of two underlying velocities $v_x$ and $v_y$. Since the Gaussian distribution for a single velocity \textit{v} is given by;
\begin{equation}
P(v|\sigma)=\displaystyle\frac {1}{\sqrt{2\pi} \sigma} e^{-\frac{v^2}{2\sigma^2}}dv
\label{equa:P.1}
\end{equation}
And we know that the tangential or 2D velocity on Earth ($v_{xy}$) is calculated by the proper motion observed i.e.
\begin{equation}
v_{xy}  =  \sqrt{v_x^2+v_y^2} \Rightarrow {v_{xy} }^2 =  {v_x^2+v_y^2}
\end{equation}
From (2) we know the probability of $v_x$ and $v_y$ is given by:\\
\begin{equation}
P(v_x|\:\sigma)=\displaystyle\frac {1}{\sqrt{2\pi} \sigma} e^{-\frac{{v_x}^2}{2\sigma^2}}d{v_x}^2
\end{equation}
\begin{equation}
P(v_y|\:\sigma)=\displaystyle\frac {1}{\sqrt{2\pi} \sigma} e^{-\frac{{v_y}^2}{2\sigma^2}}d{v_y}^2
\end{equation}
We can re-write the tangential velocity as :
\begin{equation}
P({v_x}{v_y}|\:\sigma)=\displaystyle\frac {1}{2\pi \sigma^2} e^{-\frac{(v_x^2 +v_y^2)}{2\sigma^2}}\,d{v_x}d{v_y}
\end{equation}
For the tangential velocity ($ v_{xy}$), what we actually want is the radius of the circle with x component $v_x$ and y component $v_y$. That is instead of; 
\begin{equation}
P({v_x}{v_y}|\:\sigma)d{v_x}d{v_y}
\end{equation}
We want 
\begin{equation}
P({r}{\theta}|\:\sigma)d{r}d{\theta}
\end{equation}
Via the Jacobian transformation substitution we know $d{v_{xy}} = rdrd\theta$ so we can switch to $P({r}{\theta}|\:\sigma)d{r}d{\theta}$ by substituting  $v_x^2 +v_y^2$ by $r^2$ and multiplying equation (6) by r. This gives us;
\begin{equation}
P(r\:\theta|\:\sigma) = \displaystyle\frac {r}{2\pi \sigma^2} e^{-\frac{r^2}{2\sigma^2}}\,drd\theta
\end{equation}
We also know that; 
\begin{equation}
P(r\:\theta|\:\sigma)=P(r|\:\sigma)P(\theta|\sigma)
\end{equation}
and that r is not dependant on $\theta$ so;
\begin{equation}
P(r) =\displaystyle\int_{0}^{2\pi}\ P(r,\theta)d\theta = 2\pi
\end{equation}
\begin{equation}
P(r) =\displaystyle\frac {2\pi r}{2\pi \sigma^2} e^{-\frac{r^2}{2\sigma^2}}
\end{equation}
\begin{equation}
P(r) =\displaystyle\frac {r}{\sigma^2} e^{-\frac{r^2}{2\sigma^2}}
\end{equation}
Substituting $v_{xy}$ in for r gives;
\begin{equation}
P(v_{xy} ) =\displaystyle\frac {v_{xy} }{\sigma^2} e^{-\frac{v_{xy}^2}{2\sigma^2}}
\end{equation}

\begin{table*}	
	\section{Pulsars used in the Hobbs 2D subset}
	\begin{tabular}{l c c c c c c c c c c}	
		\hline\hline
		No.&PSR&GLON&GLAT&GLON&GLAT&VZ&$\log{tc}$&Min&Dist&Velocity\\
		&&/deg&/deg&pm&pm&/km\,s$^{-1}$&/yr&pm&/kpc&2D\\
		&&&&/mas\,yr$^{-1}$&/mas\,yr$^{-1}$&&&/mas\,yr$^{-1}$&&/km\,s$^{-1}$\\
		\hline\hline
		1&B0531+21&184.63&-5.78&-14$\pm3$&-7$\pm3$&-66&3.09&14.765&2&140.0\\
		2&B0833-45&263.62&-2.77&-42.21$\pm0.09$&-17.25$\pm0.08$&-24&4.05&45.599&0.29&62.7\\
		3&B1757-24&5.29&-0.9&-1$\pm9$&-3$\pm9$&-74&4.19&3.162&5.22&78.3\\
		4&B2334+61&114.36&0.22&0$\pm13$&-10$\pm13$&-149&4.61&10.000&3.15&149.3\\
		5&B0611+22&188.86&2.4&0$\pm7$&-5$\pm6$&-49&4.95&5.000&2.08&49.3\\
		6&B1951+32&68.84&2.81&-14$\pm4$&17$\pm4$&201&5.03&22.023&2.5&261.0\\
		7&B0656+14&201.18&8.26&17$\pm3$&44$\pm3$&60&5.05&44.165&0.29&60.7\\
		8&B1830-08&23.46&0.06&-2$\pm11$&33$\pm5$&729&5.17&33.061&4.66&730.3\\
		9&B0740-28&243.85&-2.43&-13.6$\pm0.2$&-22.6$\pm0.2$&-222&5.2&26.377&2.07&258.8\\
		10&B1822-09&21.52&1.32&-12$\pm7$&9$\pm11$&38&5.37&15.000&0.88&62.6\\
		11&B0540+23&184.44&-3.32&-1$\pm8$&23$\pm8$&224&5.4&22.472&2.06&219.5\\
		12&B0114+58&126.36&-3.47&10$\pm6$&17$\pm6$&179&5.44&19.723&2.23&208.5\\
		13&J0633+1746&195.21&4.27&-33$\pm4$&178$\pm4$&135&5.53&168.680&0.16&127.9\\
		14&B0136+57&129.29&-4.06&-6$\pm5$&-20$\pm5$&-272&5.61&20.881&2.88&285.1\\
		15&B2011+38&76&2.46&-35$\pm3$&13$\pm2$&523.59&5.61&37.336&8.44&1493.8\\
		16&B1913+10&44.78&-0.66&8$\pm5$&-6$\pm$5&-178&5.62&10.000&6.27&297.2\\
		17&B2002+31&69.08&0.01&7$\pm9$&-11$\pm11$&-391&5.65&13.038&7.5&463.6\\
		18&B1838-04&27.89&0.27&11$\pm8$&-10$\pm5$&-269&5.66&14.866&5.68&400.3\\
		19&B0919+06&225.48&36.4&-64.95$\pm0.11$&61.69$\pm0.08$&282&5.7&88.484&1.2&503.4\\
		20&B1924+16&51.93&0.05&9$\pm10$&-16$\pm12$&-442&5.71&18.358&5.83&507.4\\
		21&B2148+52&97.59&-0.93&-9$\pm3$&0$\pm3$&7&5.72&9.000&4.62&197.1\\
		22&B0355+54&148.26&0.8&0.4$\pm0.3$&13.5$\pm0.4$&70&5.75&12.324&1.1&64.3\\
		23&B2351+61&116.31&-0.21&25$\pm3$&1.5$\pm2.1$&24&5.96&22.804&3.43&370.8\\
		24&B1933+16&52.51&-2.1&-6$\pm3$&-5$\pm3$&-133&5.98&7.810&5.61&207.7\\
		25&B2255+58&108.9&-0.59&-17$\pm5$&-3$\pm5$&-64&6&17.263&4.51&369.1\\
		26&B1449-64&315.8&-4.42&-18.9$\pm1$&-10.7$\pm0.9$&-105&6.02&21.719&2.08&214.2\\
		27&B1749-28&1.61&-0.96&-5$\pm6$&2$\pm6$&11&6.04&5.385&1.23&31.4\\
		28&B2224+65&108.71&6.83&185$\pm3$&19$\pm3$&166&6.05&182.428&1.86&1608.6\\
		29&B1900+06&39.89&0.33&11$\pm12$&-1.8$\pm9.1$&-72&6.14&11.146&8.44&446.0\\
		30&B1818-04&25.53&4.73&-1$\pm9$&13$\pm5$&119&6.18&13.038&1.94&119.9\\
		31&B1706-16&5.85&13.66&3$\pm13$&-2$\pm11$&-6&6.21&3.000&0.83&11.8\\
		32&B1946+35&70.77&5.03&-1.8$\pm0.6$&10.7$\pm0.6$&293&6.21&10.850&5.8&298.3\\
		33&B0402+61&144.1&7.04&40$\pm11$&54$\pm9$&539&6.23&67.201&2.12&675.4\\
		34&B1907+10&44.91&0.98&1$\pm7$&8$\pm7$&158&6.23&8.062&4.18&159.8\\
		35&B1914+09&44.63&-1.03&10$\pm6$&5$\pm6$&70&6.23&11.180&2.94&155.8\\
		36&B0458+46&160.43&3.07&-12$\pm5$&-1$\pm4$&-5&6.26&11.314&1.39&74.6\\
		37&B1904+06&40.68&-0.31&8$\pm10$&3$\pm9$&118&6.3&8.544&8.31&336.6\\
		38&B0450+55&152.69&7.54&43$\pm5$&32$\pm5$&101&6.36&53.600&0.67&170.2\\
		39&B1508+55&91.4&52.27&-16.9$\pm1.5$&96$\pm1.4$&276&6.37&97.476&0.99&457.5\\
		40&B2022+50&86.94&7.53&-11$\pm3$&19$\pm3$&209&6.37&21.954&2.34&243.5\\
		41&B0628-28&237.03&-16.75&-32$\pm3$&-34.3$\pm1.2$&-226&6.44&46.909&1.45&322.4\\
		42&B1325-43&309.95&18.43&17$\pm8$&53$\pm23$&319&6.45&54.083&1.34&343.6\\
		43&B2045-16&30.58&-33.08&42$\pm5$&-107$\pm5$&-238&6.45&114.948&0.56&305.2\\
		44&B0834+06&219.79&26.28&-43$\pm4$&30$\pm5$&83&6.47&51.039&0.65&157.3\\
		45&B2021+51&87.93&8.36&11.1$\pm0.3$&11.3$\pm0.3$&106&6.44&12.633&2&119.8\\
		\hline
	\end{tabular}
\end{table*}
\onecolumn
\begin{figure}
	\hspace{-5mm}
	\centering	
	\section{Endpoints from the evolution of the primary and secondary stars in a binary system}
		\includegraphics[scale=0.70]{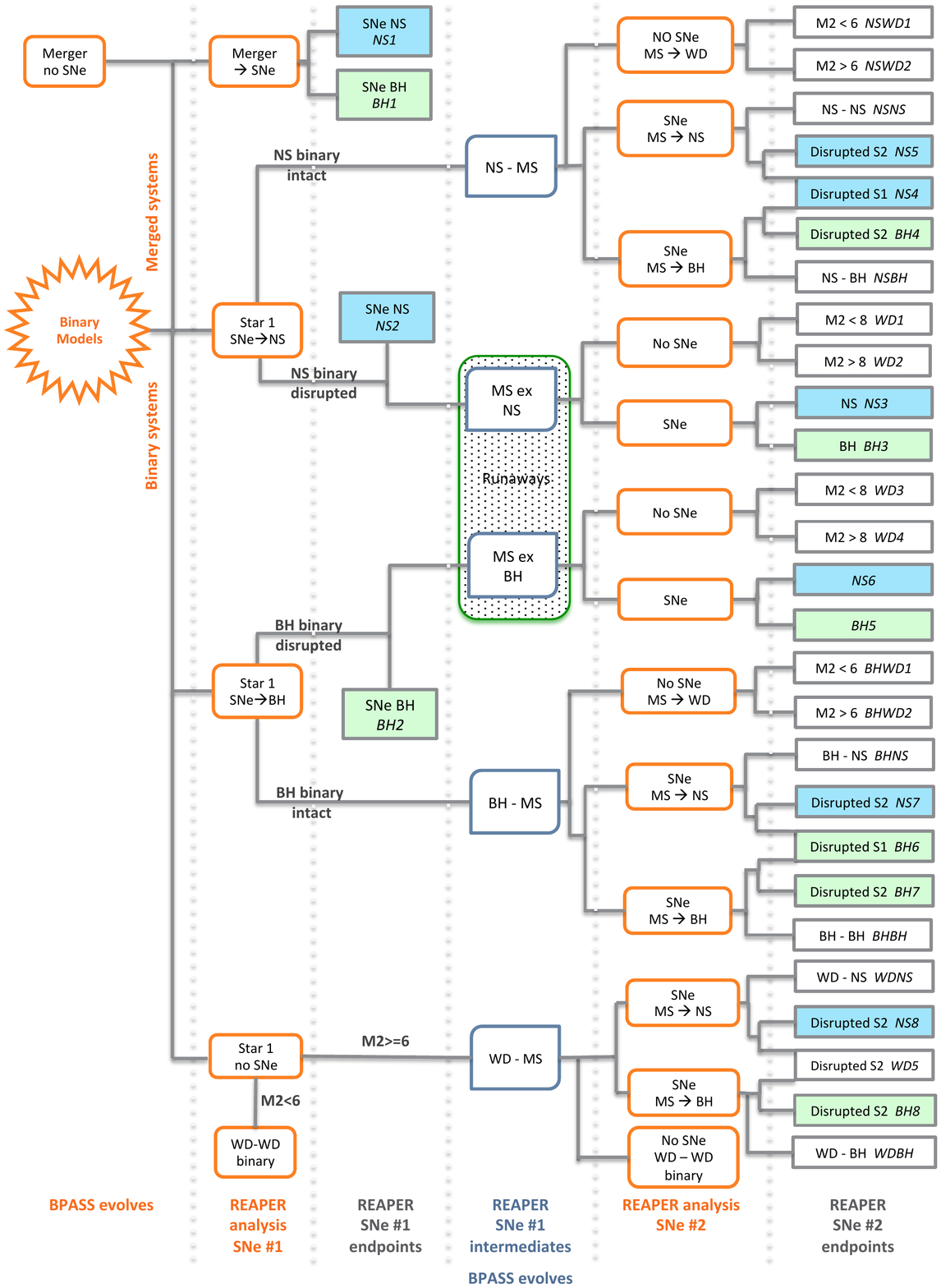}\\
	Key: 	SNe=supernova : NS=neutron star : BH=black hole : MS=main sequence : WD=white dwarf\\
\end{figure}
\end{appendices}
\end{document}